\documentclass[useAMS,usenatbib]{mn2e}
\usepackage{epsfig}
\usepackage{lscape}
\usepackage{epsfig}
\usepackage{rotating}
\usepackage{setspace}
\usepackage{pdflscape}
\usepackage{changepage}
\usepackage[dvips]{color}

\title[Discovery of wide low and very low-mass binary systems using Virtual Observatory tools]{Wide low and VLM binary systems using VO tools}
\author[M. C. G\'alvez-Ortiz et al.]{M. C. G\'alvez-Ortiz,$^{1,2}$\thanks{E-mail:mcz@cab.inta-csic.es}
E. Solano,$^{3,4}$ N. Lodieu, $^{5,6}$ and M. Aberasturi$^{7}$\\
$^{1}$ Centro de Astrobiolog\'ia (CSIC-INTA), Ctra. Ajalvir km 4, E-28850 Torrej\'on de Ardoz, Madrid, Spain.\\
$^{2}$ Centre for Astrophysics Research, Science and Technology Research Institute, University of Hertfordshire, Hatfield AL10 9AB, UK.\\
$^{3}$ Centro de Astrobiolog\'{\i}a (INTA-CSIC), Departamento de Astrof\'{\i}sica, P.O. Box 78, E-28691 Villanueva de la Ca\~{n}ada, Madrid, Spain.\\
$^{4}$ Spanish Virtual Observatory. \\
$^{5}$ Instituto de Astrof\'isica de Canarias (IAC), Calle V\'ia L\'actea s/n, E-38200 La Laguna, Tenerife, Spain.\\
$^{6}$ Departamento de Astrof\'isica, Universidad de La Laguna (ULL), E-38205 La Laguna, Tenerife, Spain IAC.\\
$^{7}$ ESA--ESAC, SOC. P.O. Box 78, E-28691 Villanueva de la Ca\~nada, Madrid, Spain.}

\begin{document}

\date{Accepted 1988 December 15. Received 1988 December 14; in original form 1988 October 11}

\pagerange{\pageref{firstpage}--\pageref{lastpage}} \pubyear{201}

\maketitle

\label{firstpage}

\begin{abstract}

The frequency of multiple systems and their properties are key constraints of stellar formation and 
 evolution. Formation mechanisms of very low-mass (VLM) objects are still under 
 considerable debate and an  
 accurate assessment of their multiplicity and orbital properties are essential for
  constraining current theoretical models.

Taking advantage of the Virtual Observatory capabilities, we 
 looked for comoving low and VLM binary (or multiple) systems using the Large 
 Area Survey of the UKIDSS LAS DR10, SDSS DR9, and the 2MASS Catalogues. Other catalogues 
 ($WISE$, GLIMPSE, SuperCosmos ...) were used to derive the physical parameters of the systems.

 We report the identification of 36 low and VLM ($\sim$M0-L0 spectral types) candidates 
 to binary/multiple system (separations between 200 and 92000 AU), 
 whose physical association is confirmed through common proper motion,
 distance and low probability of chance alignment. This new system list notably increases
  the previous sampling in their mass-separation parameter space ($\sim$100).
  We have also found 50 low-mass objects that we can classify as $\sim$L0-T2
 according to their photometric information. Only one
 of these objects presents a common proper motion high-mass companion.

 Although we could not constrain the age of the majority of the candidates, probably
 most of them  are still bound except four that may be under disruption processes.
 We suggest that our sample could be divided in two populations:
 one tightly bound wide VLM systems that are expected to last more than 10 Gyr,
 and other formed by weak bound wide VLM systems that will dissipate within a few Gyrs.

\end{abstract}


\section{Introduction}

Binary and multiple stars have long provided an empirical effective method of
  testing stellar formation and evolution theories. 

 Our understanding of formation and evolution processes in very 
 low-mass stellar (VLM) and substellar objects is still not clear. 
 Here we define VLM stars as fully convective stars with masses under 0.3 M$\sun$
 (which typically corresponds to spectral types M3-M4) and above the substellar limit 
 (0.075 M$\sun$; Mart\'in 2000). Ultracool dwarfs (UCDs) refers to objects 
 with spectral type of M7 or later corresponding to $T_{eff}$ $\leq$ 2500 K,
 the spectra of which are dominated by molecular bands and dust
  (e.g. Knapp et al. 2004; Burgasser, Burrows \& Kirkpatrick 2006). 

 Whether these cool objects form in a similar manner to higher mass stars or require
 additional or different processes is currently under debate 
 (e.g., Burgasser et al. 2007; Luhman et al. 2007; Whitworth et al. 2007; Luhman 2012; 
 Chabrier et al. 2014). 
 Since multiplicity properties such as binary fraction, period and mass-ratio distribution
 or separation, provide important constraints on 
 star formation and dynamical evolution (Burgasser et al. 2007; Goodwin \& Whitworth 2007), 
 it is of considerable interest to determine whether the properties 
 of VLM binaries differ from those of more massive stars (see, for instance, Duquennoy \& Mayor 1991 and 
 Fischer \& Marcy 1992).

 Many works have estimated binary fraction and the mass-ratio distribution.
 Burgasser et al. (2007) presented an extensive review on the multiplicity of VLM objects, reporting 
 important differences between them and the more massive stars.
 In particular, VLM objects show a lower multiplicity ratio, 20-25\% (Kraus \& Hillenbrand 2012), 
 compared to $\approx$80\% for B stars (Kouwenhoven et al. 2005), $\approx$65\% for G stars 
 (Duquennoy \& Mayor 1991) and $\approx$40\% for M2-M5 dwarfs (Fischer \& Marcy 1992), 
 and the mass-ratio distribution strongly peaks at unity whereas it is more evenly 
 distributed for more massive stars. Also, the separation between 
 components is significantly smaller for VLM objects. Typical separation is $\sim$4 AU
 (e.g. Close et al. 2003; Burgasser et al. 2007;  Kraus \& Hillenbrand 2012), from
  statistical samples with separation over 3 AU due to the resolving power of imaging programmes.
 If works with samples with separation under 3 AU are included the total binary fractions for VLM stars
 and brown dwarfs (BDs) could range between 2-50\% (Bardalez Gagliuffi et al. 2014).
 These differences can be due either to a continuous formation mass-dependent trend 
 or to differences in the formation mechanism of the VLM objects.

 Wide ($>$ 100 AU) binary companions are relatively common for high-mass stars.
  Raghavan et al. (2010) found that $\sim$25\% of solar-type stars have a companion 
 with separations wider than 100 AU and $\sim$11\% wider than 1000 AU. 
 Tokovinin \& L\'epine (2012) estimates that at least 4.4\% of the solar-type stars 
 have a companion at more than 2000 AU. Nevertheless, different projects dedicated to 
 find wide VLM binaries concluded that they are rare. Allen (2007), via statistical 
 investigation utilizing a Bayesian algorithm, found that only 2.3\%
 of VLM objects have a companion in the 40-1000 AU range.
 Burgasser et al. (2009) estimated a fraction of VLM wide multiples in
 the field of no more than 1\%-2\%. For a recent review on stellar multiplicity,
 see Duch$\hat{e}$ne \& Kraus (2013).

 Searches for VLM companions to M dwarfs at large separations ($>$ 100 AU) 
 remain incomplete. The discovery of VLM binaries with separations 
 of thousands AU (e.g., Artigau et al. 2007, Caballero 2007a,b, Radigan et al. 2009) point to the 
 existence of a population of such ultra-wide systems. 
 Dhital et al. (2010), presented the Sloan Low-mass Wide Pairs of Kinematically 
 Equivalent Stars (SLoWPoKES) catalogue with 1342 very-wide (projected separation
  over 500 AU) low-mass (at least one mid-K to mid-M dwarf component)
  common proper motion pairs identified from astrometry, photometry, and proper motions in the 
 Sloan Digital Sky Survey (SDSS).
 They found that the wide binary frequency was at least of $1.1\%$ in the 
 spectral types of the catalogue with many pairs so weakly bound ($\approx$10$^{33}$J 
 for the weakest) that overcome the 
 limits of previous empirical limits (Close et al. 2003; Burgasser et al. 2007). The
 catalogue also presents a bimodality in binary separation 
 (as marked by previous works, e.g. Kouwenhoven et al. 2010), suggesting the presence of 
 two populations, one "old" and tightly bound, with binding energies of formation to survive
 the age of the Galaxy and other "young" weakly bound systems, recently formed and that 
 will not survive more than 1-2 Gyr. 
 Janson et al. (2012) presented the results of an extensive high-resolution
 imaging survey of M-dwarf multiplicity, where they found a multiplicity fraction of 27$\pm$3$\%$
 for M-dwarfs within the AstraLux detection range of 0.08''--6'', semimajor axes in 3-227 AU range
 at a median distance of 30 pc. They concluded that their results indicate a common
 formation mechanism between stars and BDs.
 Also, Baron et al. (2015) reported the discovery of 14 VLM binary systems formed by
 mid-M to mid-L dwarf companions with separations of 250-7500 AU.

 Despite the relevant contribution of the mentioned works in the field of
 VLM wide binaries, the number of known pairs formed by mid to late M (or cooler) dwarfs
 is still small.
 
 A compilation from the articles previously reported gives $\sim$100 wide systems
 formed by M or later components and $\sim$40 with components with M5
 or later spectral types.
 During the writing of this paper, Dhital et al. (2015) present the second part of
 SLoWPoKES, that contains a significant number of binaries at the late-M spectral types.
 They identified 944 sdM + sdM binary candidates and 141 in which both components are VLMs,
 although spectroscopic data are needed to confirm them.

Recently also higher order multiplicity has also been explored as key
 to  constrain the star formation processes.
 Faherty et al. (2010) found that the frequency of tight resolved binaries
 in wide systems ($>$ 100 AU) that contains an UCD ($>$ M6) is at least twice (50\%)
  the frequency for wide isolated field UCDs (10-20\%).
 They obtained values of 3:5 and 1:4 ratios of triples
 and quadruples to binaries respectively, in comparison with 1:4 and 1:26
 ratios from Reid \& Hawley (2005) M dwarfs sample.
 Law et al. (2010) studying a sample of 36 extremely wide (600--6500 AU) M1-M5 dwarf binaries
 obtained a  bias-corrected, high-order-multiple fraction of 45\%, with a
 quadruple incidence inferior to 5\%.
 The authors reported an increase in the high-order multiple fraction for the widest
 targets, reaching 21\% for systems with separations up to 2000 AU and 77\%
 for systems with separations over 4000 AU. Also, Allen et al. (2012) estimated
  a wide tertiary fraction of 19.5\% from a multi-epoch search 
 for wide ($>$ 200 AU) low-mass tertiary companions of a volume-limited sample of
  118 known spectroscopic binaries within 30 pc of the Sun.

 These wide binaries, with large separations and low binding energies, have a strong impact on the
 proposed formation theories. In particular, it challenges the ejection model
 (Reipurth \& Clarke 2001; Bate \& Bonnell 2005) since such fragile systems are not expected to survive the ejection
 process from their birth environments. They also raise some concerns on the disc fragmentation scenario
 (Padoan \& Nordlund 2002) as such wide systems would require the existence of discs of unreasonable mass and
 size. The photo-evaporation mechanism (Whitworth \& Zinnecker 2004), on the other hand, demands the presence of nearby 
massive young stars,
 and therefore cannot be a universal mechanism for VLM multiple formation (Burgasser et al. 2007). Also the massive
 stars would probably disrupt the wide VLM multiple system.

Kouwenhoven et al. (2010) and (2011) remarked that wide binary systems, with separations larger
  that 1000 AU, could not have been formed as primordial binaries in star clusters
 since their orbital separation would be comparable to the size of a typical embedded cluster. They proposed,
 based on N-body simulations, that these binary systems were formed during
 the star cluster dissolution process, since escaping stars would have very similar velocities, easing the
formation of wide systems. Based on the relatively high percentage of binaries found (up to 30 \%) they also predicted
a high frequency of triple and quadruple systems.

 The discovery of wide binaries in large regions of the sky typically requires managing huge volumes of data coming
  from different astronomical archives and services which, once discovered and gathered, need 
 to be cross-matched and filtered following a number of photometric and kinematic criteria.
  The drawbacks associated with this type of analysis can be overcome if the work is done
  in the framework of the Virtual Observatory (VO\footnote{http://www.ivoa.net}), an 
 international initiative whose main goal is to guarantee an easy access and analysis to the
  astronomical data distributed worldwide. 

Making use of VO tools, we searched for wide binaries exploring the limits
 in distance and binding energy. We have discovered new 36 low-mass and 
 VLM multiple systems, with averaged projected 
 separation between $\sim$200 and $\sim$92000 AU, using the
 Two Micron All Sky Survey (2MASS) Point Source Catalogue (PSC; Skrutskie et al. 2006),
 the Sloan Digital Sky Survey (SDSS) Data Release 9 (DR9) Photoprimary Catalogue 
 (Adelman-McCarthy et al. 2009), the Large Area Survey of the United Kingdom 
 Infrared Telescope Infrared Deep Sky Survey Data Release 10 (UKIDSS LAS DR10; 
 Lawrence et al. 2007), the Wide-field Infrared Survey Explorer ($WISE$; Wright et al. 2010),
  and The Galactic Legacy Infrared Mid-Plane Survey Extraordinaire 
 (GLIMPSE\footnote{http://www.astro.wisc.edu/glimpse}) data bases. 
 Also, we have found 50 low-mass objects with $\sim$L0-T2 photometric spectral types. Only one
 of these objects presents a common proper motion high-mass companion, inside our
 research limits.  
  The methodology is outlined in Section 2, while Section 3 describes the 
 proper motions. Sections 4 and 5 present the analysis of general properties of the new systems
 and of the L dwarfs respectively. Section 6 presents the search of high-mass companions.
 Section 7 presents the discussion of results and the summary and conclusions are given in Section 8.

\section{Search methodology}

 We searched for common proper motion objects 
 with colours consistent with spectral types later than M0,
 using STILTS\footnote{http://www.star.bris.ac.uk/$\sim$mbt/stilts/} 
 to build a workflow consisting of the following steps:

\begin{itemize}
\item Cross-matching: 
 We performed a cross-match between the 2MASS and the SDSS DR9
 Catalogues in the whole area of the sky covered by SDSS ($\approx$14555 squares degrees).
 We used the command $tskymatch2$ with the option $best$. This makes a crossmatch of two tables 
 based on the proximity of sky positions. The best pairs were selected in a way which treats the two 
 tables symmetrically. Any input row which appears in one result pair was disqualified from appearing 
 in any other result pair, so each row from both input tables appears in at most one row in the result.
 We used a matching radius of 10 $\arcsec$ to ensure that objects with high proper motion are not
 left out and, at the same time, keep the management of false alarms tractable.
 Considering the maximum possible time difference between 2MASS and SDSS observations
  (12 years, 1997-2009), a separation of 10 $\arcsec$ implies that we are able to find all 
 objects with proper motions less than 0.8 $\arcsec$/yr. Faherty et al. (2009), in their 
 kinematic study of late-type dwarfs, concluded that only $\approx$ 10$\%$ of M dwarf have 
 proper motions higher than 0.8 $\arcsec$/yr, which confirms the high level of completeness 
 of our criterion.
 Only the pair 2MASS-SDSS with the minimum separation was considered. We also checked that this pair 
 coincides with the closest SDSS-2MASS pair. Finally, given the poorer 2MASS spatial 
 resolution compared to SDSS, we required that each 2MASS source matched a unique SDSS source within 6 $\arcsec$.

 Using similar criteria, we also required the presence of a UKIDSS LAS (DR10) counterpart at less than
 20 $\arcsec$ from the 2MASS source fulfilling a $ppErrbit$ quality flag smaller than 256 and class 
 star parameter $mergedClass$=-2, -1. 
 In those sky regions not covered by UKIDSS, AllWISE\footnote{http://cdsarc.u-strasbg.fr/viz-bin/Cat?II/328} 
 was used instead.
 To ensure a real displacement of the source between the 2MASS and UKIDSS ($WISE$) images we required a 
 separation 2MASS-UKIDSS (2MASS-$WISE$) larger than 0.7 $\arcsec$ (the 90 percentile in separation using all objects 
 in the image is typically $\approx$ 0.6 $\arcsec$).

\item Filtering: The sources obtained from the cross-match were filtered using the following criteria:

\begin{itemize}
\item Xflg=0, Aflg=0, to avoid sources flagged in 2MASS as minor planets or contaminated from nearby extended 
sources.
\item $J$$\leqslant$ 17, to get rid of too faint objects which typically have associated large 
 photometric uncertainties.

\item Qflg($J$) $\neq$ "U", to discard sources with upper limits in the $J$ band.

\end{itemize}

\item Photometric cuts: we kept objects fulfilling the following colour criteria:

\begin{itemize}

\item Objects with M spectral type (West et al. 2011): \\
 (0.48 $\leq$ $(r-i)$ $\leq$ 2.90) \textbar \textbar (0.27 $\leq$ $(i-z)$ $\leq$ 1.89)  \\
 \& $r$ $<$ 22.2 \& $i$ $<$ 21.3 \& $z$ $<$ 20.5 \\

\item Objects with L-T spectral types (Schmidt et al. 2010): \\
$(z-J)$ $\geq$ 2.0 \& $z$ $<$ 20.5 \& ($(i-z)$ $\geq$ 1.7 \textbar \textbar$ $ $i$ $>$ 21.3) \& ($(i-J)$ $\geq$ 3.1 \textbar \textbar$ $ $i$ $>$ 21.3)

\end{itemize}

\item Proper motion cuts:
 we required that the sign of the proper motion calculated using 2MASS-UKIDSS
 (2MASS-$WISE$ in those regions not covered by UKIDSS) both 
 in RA and DEC was the same of the proper motion derived using 2MASS-SDSS. 
 This check in the two components allowed us to discard a significant number of false companions having a proper
 motion with a similar modulus but moving in different direction in the space.

\end{itemize}

 We selected objects fulfilling the above criteria and differing $<$ 30$\%$ in each component of 
 the proper motion ($\mu_{\alpha}*cos_{\delta}$, $\mu_{\delta}$) (see Sect. 3) 
 and $<$ 20$\%$ in distance (see Sect. 4). Although the common proper motion criterion may look
 rather conservative compared to other criteria found in the literature 
 (typically $\Delta\mu / \mu$ $<$ $\sim$ 0.2: see, for instance, Dupuy \& Liu 2012),
 we note that the great majority of the common proper motion pairs given in Tables~\ref{tab:obs}
 and \ref{tab:obs2} fulfill this relation and even $\sim$40\% 
 have $\Delta\mu / \mu$ $<$ 0.1.
 We also imposed a condition on separations, rejecting system with
 angular separation smaller than 1 $\arcsec$ and larger than 100,000 AU.
 Finally, the remaining candidates were 
 visually inspected to eliminate artifacts and spurious matches.
 The search of companions was performed in steps, where the sky was divided first
 in one square degree fields in which all the criteria were applied and with
 the final output of total candidates. We therefore have not kept the intermediate
 numbers, objects per field, although it will give a rough estimation 
 that the search was performed around $\sim$ 100,000 objects in all sky covered by SDSS.

 This way we ended up with 39 candidate pairs and three objects candidates
 to form a triple system (Table~\ref{tab:obs}) in the area common between 2MASS (PSC), SDSS (DR9) and
 UKIDSS LAS (DR10). The 40 systems are form by components with $\sim$M0-L0 photometric spectral types.
 Six of these candidates were previously identified in the 
 Washington Double Star Catalog (WDS; Manson et al. 2001) and one in the recently released SLoWPoKES-II.
 All of them were removed from our list.
 In those regions not covered by UKIDSS we identified six candidates to 
 binary/multiple systems in the area in common between the 2MASS, SDSS, and
 $WISE$. One remaining candidate was found as byproduct using
  2MASS, $WISE$ and GLIMPSE positions. These seven systems have
 components with $\sim$M0.5-L0 photometric spectral types.
  One of them is candidate to a triple system (Table~\ref{tab:obs2}).
 From this list, four objects previously identified in the WDS catalogue were also removed.
 No candidate to binary/multiple system were found with spectral types later than L0.
 As part of our analysis we also identified
  50 new L-T candidates according to our photometric cuts, with no low-mass proper motion 
 companion inside the settled constraints (Table~\ref{tab:obsL}).

 Finally, we ended up with 36 pair/multiple identified as potential common proper motion objects.
 Additional eleven pairs were previously reported in SLoWPoKES II (1) and in the
 WDS catalogue (10). We kept them in our list to improve their physical parameters.
 In particular, spectral types were calculated for seven of them for the first time.
 None of our 36 systems were found in Slowpokes I,
 Faherty et al. (2009; 2010), Caballero (2009), Radigan et al. (2009), Zhao et al. (2011), 
 Luhman et al. (2012), Mu$\breve{z}$i\'c et al. (2012), Deacon et al. (2014) or Baron et al. (2015).

\begin{figure}
   \centering
   \includegraphics[width=8.5cm,clip]{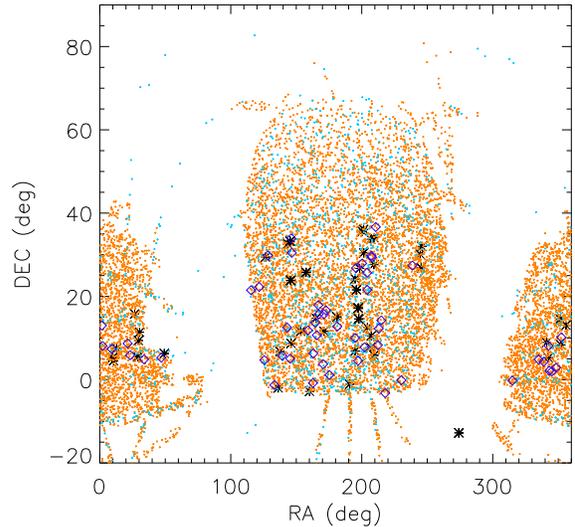}
      \caption{Spatial distribution of the binary (black asterisks) and L dwarf
 (dark blue diamonds) candidates compared to that of SLoWPoKES binaries
 (orange small crosses) and SLoWPoKES VLM binaries (light blue small crosses).
 }
         \label{fig:radeca}
   \end{figure}

 Figure~\ref{fig:radeca} displays the spatial distribution of our
 candidate list compared to SLoWPoKES.

\section{Common pairs}

\subsection{Proper motions}
 
 Proper motions were calculated from the 
 differences in position between the 2MASS, SDSS and UKIDSS epochs for a list of
 targets that we call group A and between the 2MASS, SDSS and $WISE$ for a list that we
 call group B. Whenever possible, other catalogues (the first  and the second Palomar 
 Observatory Sky Survey (POSS-I and POSS-II), GLIMPSE, etc) 
 were used to add new epochs and improve the calculated values.
 The proper motion of each component was calculated in the standard way 
 by single least squares fit to all the available positions.

\begin{figure*}
   \centering
   \includegraphics[width=17.5cm,clip]{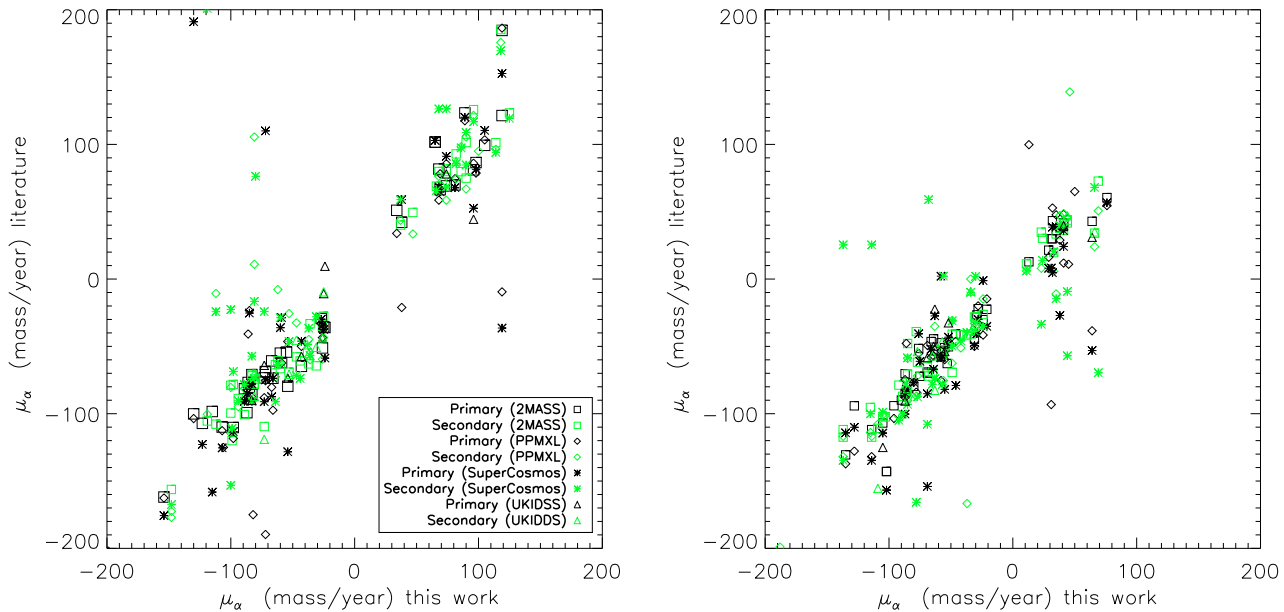}
      \caption{Comparison of proper motion measurements with literature sources.
 Primaries and secondaries are in black and green character, respectively.
 }
         \label{fig:propercompa}
   \end{figure*}

 The positions for each system component in 2MASS, SDSS and UKIDSS/$WISE$ catalogues are given in
  Tables~\ref{tab:obs} and~\ref{tab:obs2}.  
 Final proper motions are given in last two columns of the Tables,
 in [$\mu_{\alpha}*cos_{\delta}$, $\mu_{\delta}$] format for proper motion components in column 8 and 
 total proper motion in column 9 with the error from the fit rounded to the nearest whole number in parenthesis,
  both in mas yr$^{-1}$.
 Figure~\ref{fig:propercompa} shows a comparison between our determination of proper motion
 and that of some literature sources.
 We see how good agreement is reached in all cases with 2MASS and UKIDSS measurements although some 
 dispersion is found with the SuperCosmos Sky Survey (Hambly et al. 2001a, 2001b, 2001c) and Position and
  Proper Motion Extended-L catalogue (PPMXL; Roeser et al. 2010)
  measurements. The values we obtained for the SLoWPoKES II and WDS objects are in agreement with 
 the correspondent catalogue results, except for 2MASS J13104398+1434338 / 2MASS J13104431+1434326 pair 
 for which no proper motion is given in the WDS catalogue.
 The highest proper motions found in our list of objects is $\approx$ 0.4 $\arcsec$/yr.

 As our nearest pair is at $\sim$26 pc (Table~\ref{tab:parab}), we assume that the error produced
  by parallax in the proper motion measurements is negligible.

\subsection{Astrometric confirmation}


 To further assess the reliability of the proper motions of the binary systems,
  we measured the variation of the angular separation, $\rho$, and orientation or parallactic angle,
 $\theta$, defined as $\theta$=$atan (\delta(RA) / \delta(DEC)$), between the components of each pair. 
 Five epochs (POSS-I, POSS-II, 2MASS, SDSS, UKIDSS) or a span of $\sim$50 years
 were available for 31 pairs, four epochs (POSS-II, 2MASS, SDSS, UKIDSS) or about 20 years of
 time base-line for 6 pairs and three epochs (2MASS, SDSS, UKIDSS) or a coverage of $\sim$10 years
  for 3 pairs.
 Note that the 36 systems are formed by 34 binaries and 2 triples, what 
 gives a total of 40 pairs considering the components of the two triples in pairs.
 If the pairs are real common proper motion objects, $\rho$ and $\theta$ should remain constant
 within the errors. To check this, we compared the mean value of $\rho$ and $\theta$ with the values
  that these parameters should have if the second component of the system was a fixed, background star.
 We assumed common proper motion provided that the difference in $\rho$ and $\theta$ is higher than
 3$\sigma$ for time baselines of 50 years, 2$\sigma$ for epochs separated 20 years and 1$\sigma$ for
  baselines of 10 years. All systems fulfilled these criteria.

\subsection{Chance alignments}

 The probability of random alignment is the probability that, for chance, in a search 
 like ours, we find physically unrelated stars having the same proper motion and 
 distance within our uncertainties. The probability is calculated by multiplying 
 the probability of finding two stars with the same proper motion by the probability 
 of finding two stars at the same distance.

 We looked for PPMXL sources with the
 same proper motion within the assumed uncertainties (difference of less
 than 30\% in both components) in a cone of 10$^{\prime}$ radius (slightly larger
 than the largest separation between components) centred in every
 candidate system. We found that the probability of a single star having a common proper
 motion occurring by chance is $\sim$ 0.15\%.

 In a second step, we used the space density of M dwarfs ($\approx$ 5.4x10$^{-2}$ pc$^{-3}$, 
 from Caballero et al. 2008 and reference therein), 
 to estimate the likelihood of a close proximity in space by chance. Taking into account the
 distances and the separations of our system, we calculated a volume for each target. 
 The probability of sharing by chance the same volume of space, calculated as $density*volume$ or,
  equivalently, assuming Poisson statistics ($P = 1 - e^{-\rho*V} $; e.g. McElwain \&
 Burgasser 2006), ranges from 10$^{-1}$ to 10$^{-6}$.

  The combination of these two factors indicates that the
 probability of a single star having a chance alignment in
 space and motion at the level of our uncertainties ranges from 10$^{-3}$ to 10$^{-9}$.
 To get the chance alignment probability for the sample we need to sum
 the individual probabilities (1.2x10$^{-3}$), and then multiply by
 the number of objects searched for companions (100,000 aprox.) 
 divided by the number of objects found (96). We obtain that there might be between 
 1 and 2 chance alignment in our sample of 47 systems.
  This indicates that the majority of our candidates form physical pairs. However, we
  note that the wider the binary, more likely to be chance alignments because the
  probability of chance alignment goes up with separation.

\section{Properties of the new systems}

\subsection{Kinematics}

 True space velocities are better indicators of an object's kinematics than apparent angular motions.
 But we only have measurements of radial velocity for two candidates,
 2MASS J13570535+3403459 and 2MASS J09332493+3232033, with values of 64.2 and -12.9 km$^{-1}$ 
 respectively (errors between 5-10 km$^{-1}$) from West et al. (2008). 
 They calculated the Galactic space-velocity components $(U, V, W)$ using photometric distance 
 values of 179 and 180 pc (quite similar to the ones obtained here), 
 obtaining (-32.5, -5.0, 54.0) and (8.4, 1.9, -5.4).
 The $(U, V, W)$ coordinates of 2MASS J13570535+3403459 lay in the region of young disc population
 defined by Eggen (1984a,b, 1989), in the Ursa Major group area (500 Myr; 
 King et al. 2003).
 For 2MASS J09332493+3232033, $(U,V)$ coordinates lay also in the young disc area, but
 $W$ component is situated in the field. The $W$ component is always affected by larger 
 dispersion, so the target may still belong to the young disc area.
 West et al. (2008) calculated vertical distance from the Galactic plane of 146.2 and 188.0 pc 
 that correspond to the thin disc (e.g. Pauli et al. 2006).

\begin{figure}
   \centering
   \includegraphics[width=8.5cm,clip]{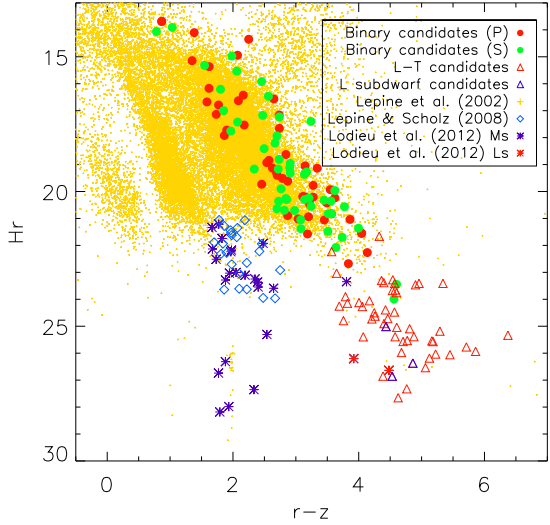}
      \caption{Reduced proper motion diagram.
 Small orange crosses represent all sources in L\'epine et al. (2002) with counterpart in 
 the SDSS DR7 data base (99\% complete for stars with proper motions 0.5''yr$^{-1}$$<\mu<$2.0''yr$^{-1}$
 down to $R$=19), blue open diamonds are known late M subdwarfs
 from L\'epine \& Scholz (2008), purple and red asterisks are M and L subdwarfs
 from Lodieu et at. (2012) respectively. Filled dots are our candidates to binary/multiple systems,
  red for primaries and green for secondaries, red triangles are the L-T candidates
 and blue triangles represent three L subdwarfs candidates.
              }
         \label{fig:hr}
   \end{figure}

 For the rest of the candidates, 
 we just made the kinematic study using reduced proper motions.
 Since subdwarfs are old and tend to exhibit halo or thick disc kinematics (Gizis 1997),
 the reduced proper motion diagram represents a useful tool to separate dwarfs
 from subdwarfs and white dwarfs objects (Jones 1972, Evans 1992, Salim \& Gould 2002,
  L\'epine \& Shara 2005; Burgasser et al. 2007, Lodieu et al. 2012).
 We have used the diagram of reduced proper motion in $r$ magnitude versus $(r-z)$ 
 SDSS colour as in Lodieu et al. (2012).
  From this diagram, Figure~\ref{fig:hr}, we can confirm that individual components 
  of the systems are consistent with solar-type dwarfs of the disc population.

\subsection{Properties from photometry}

\subsubsection{Distances}

 Photometric distances were calculated using the colour-absolute magnitude relations given
  in Bochanski et al. (2010; Table 4) for M0-M9 dwarfs. We used the $M_r-(r-z)$ relationship
  whenever possible and $M_r-(r-i)$ or $M_r-(i-z)$ otherwise. 
 The associated uncertainties are $\sigma$M = 0.394 mag, 0.403 and 0.481 mag
 (Bochanski et al. 2010; Table 4) which translate into errors of $\sim$20$\%$ in distances.
 Since the pair 2MASS J18163409-1246310 and 2MASS J18163485-1246421
 do not have an SDSS counterpart, we used the $M_J$-spectral type relation in Hawley et al. (2002)
 to estimate their distances. The error in distance associated with this calibration is not larger 
 than 25$\%$.  The calculated values are given in column 7 of 
 Tables~\ref{tab:para} and~\ref{tab:parab}.

\subsubsection{Establishing these objects
are dwarfs}

 We used L\'epine \& Gaidos (2011) photometric criterion to separate M dwarfs from giants
 and verify the dwarf nature obtained from kinematics: \\

\begin{equation}
M_{V} > 2.2 (V-J) -2.0 \\
\end{equation}

 where $V$ magnitude was calculated for our objects using the following transformation between 
 the SDSS and Johnson photometric systems\footnote{https://www.sdss3.org/dr8/algorithms/sdssUBVRITransform.php}.

\begin{equation}
V = g - 0.5784 (g - r) - 0.0038 \\
\end{equation}

 All our M spectral type candidates fulfilled the dwarf condition. 
 Since 2MASS J18163409-1246310 and 2MASS J18163485-1246421
 do not have SDSS colours, we used the equations (9)-(13) 
 of L\'epine \& Gaidos (2011) instead to confirm their dwarf nature.

\subsubsection{Effective temperatures, surface gravities and masses.}

 Effective temperatures and surface gravities were computed using
 VOSA\footnote{http://svo2.cab.inta-csic.es/theory/vosa/} (Bayo et al. 2008), assuming solar metallicity
 (as expected from dwarfs in the thin disc). 
 VOSA is a VO tool designed to determine physical parameters 
 from comparison of observed photometry
 gathering from different services (2MASS, SDSS, UKIDSS, $WISE$,...) with different collections of
 theoretical models. In particular, we used the BT-Settl models (Allard, Homeier \& Freytag 2011, 2012)
 and the two methods of comparison with models that VOSA offers, namely, a $\chi^{2}$ minimization, that 
 returns the “overall best-fitting model” and a Bayes analysis giving information on the marginalized probability
 distribution of the individual parameters in the fit. In all cases, the $T_{eff}$ obtained with the best
 $\chi^{2}$ fit has a normalized probability larger than 0.8. This was the condition we imposed to consider 
 the derived effective temperature reliable.

  We limited the range of $log~g$ in the VOSA fitting to $log~g$=4.5-6.5, to include the
 typical values for M dwarfs with ages of 0.1-10 Gyr (Jones et al. 1996) plus 0.5 dex of error. 
 VOSA also provides mass estimations, using BT-Settl models (Allard et al. 2011, 2012). 
 An average uncertainty of $\approx$ 0.015 M$_{\odot}$ was assumed.

 We also calculated masses from the $Mr$, $Mi$ and $Mz$ absolute 
 magnitudes given in Table~5 of Kraus \& Hillenbrand (2007). 
 They adopted effective temperatures from models of  Luhman (1999) for spectral types $>$M0
 and then combined these Teff values with the 500 Myr isochrones 
 of Baraffe et al. (1998) to estimate stellar masses. 
 Since 2MASS J18163409-1246310 and 2MASS J18163485-1246421 lack SDSS colours, 
 we used the relation from $J$ magnitude. 
 Since theoretical models can underpredict masses (e.g., Hillenbrand \& White 2004; L\'opez-Morales 2007),
 they increased the masses of M1 stars by 5\%, M2 stars by 10\%, and later type stars by 20\%, getting mass
 values more consistent with the observations. See Kraus \& Hillenbrand (2007) for further details.
 VOSA masses are in average 0.06 M$_{\odot}$ smaller than masses from 
 Kraus \& Hilenbrand (2007). In following calculations, we used the masses derived from 
 Kraus \& Hilenbrand (2007) calibration when VOSA values were not available.

\begin{figure*}
   \centering
   \includegraphics[width=12.5cm,clip]{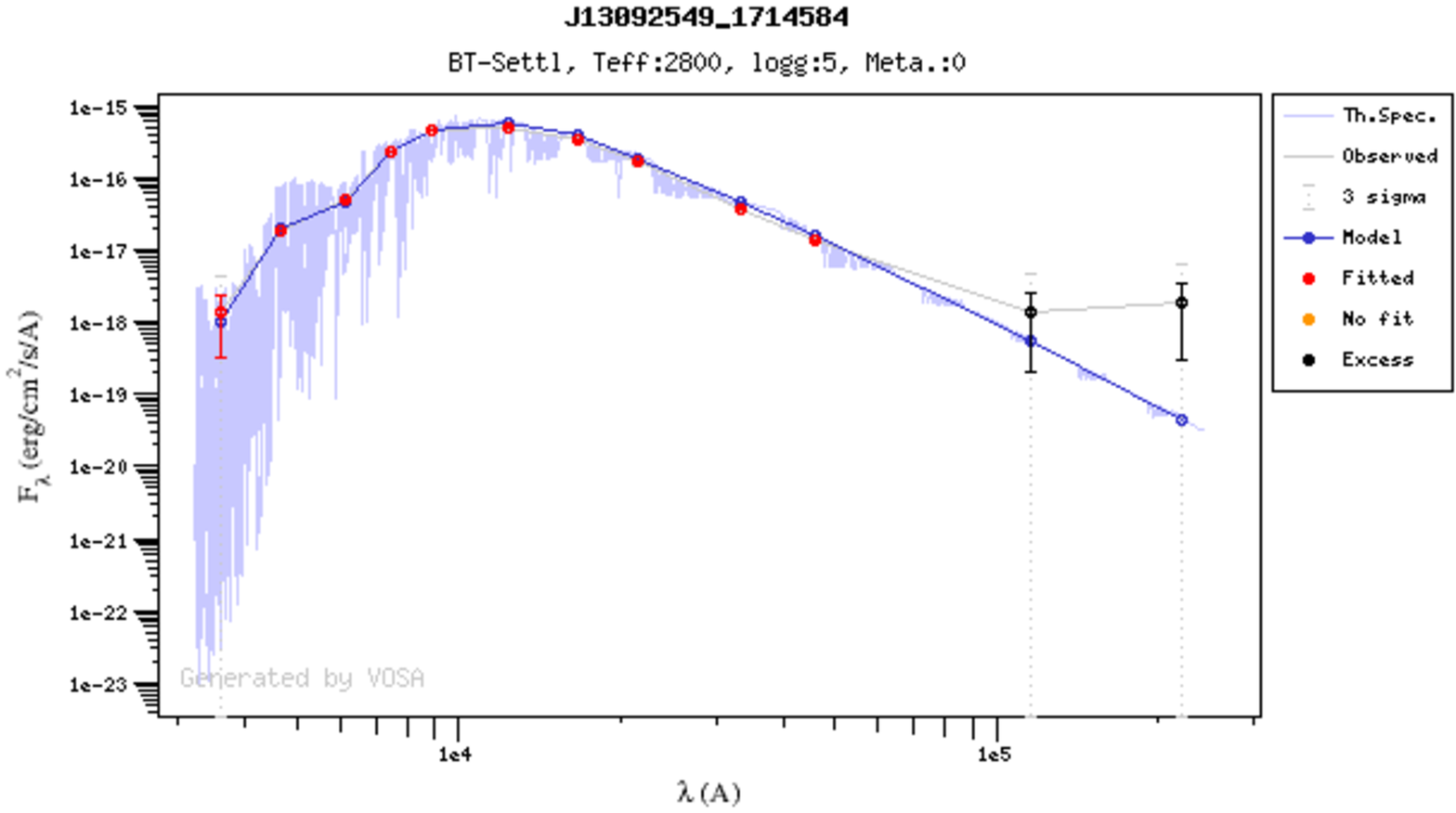}
   \includegraphics[width=12.5cm,clip]{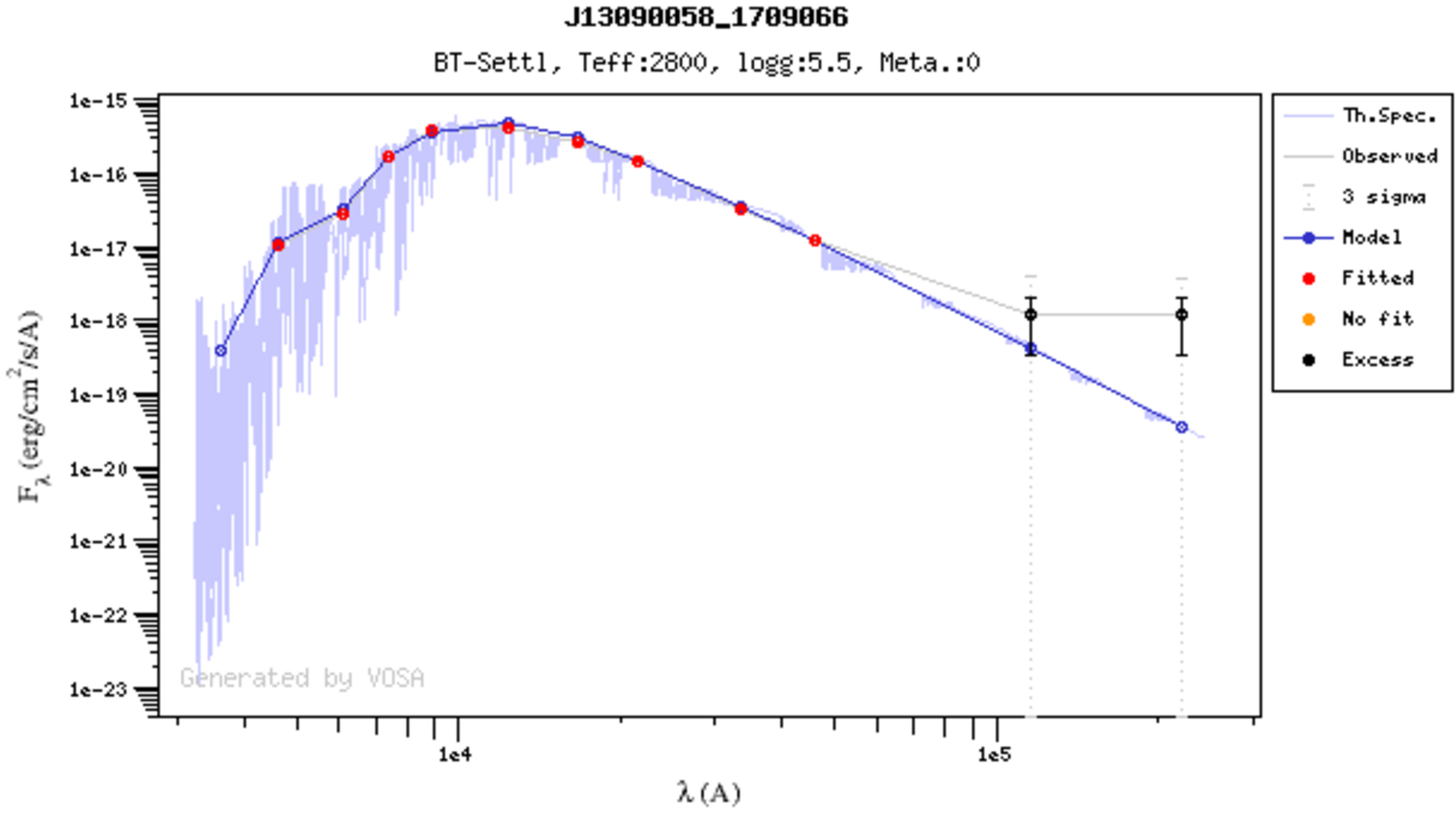}\\
      \caption{
   SED fitting plot generated by VOSA for one of the binary systems. Temperature,
   $log~g$ and metallicity of the best fit model are given.
  Blue line and dots represent theoretical spectra and photometry, 
 respectively while red dots, joined by a grey line, represent observed photometry 
 (SDSS, 2MASS, UKIDSS and $WISE$). Orange dots are photometric points that were not
 included in the fitting process. See Sect. 4.2.3 for details.
              }
         \label{fig:fig2}
   \end{figure*}

  The calculated $T_{eff}$, $log~g$ and masses are given in Tables~\ref{tab:para} and~\ref{tab:parab}.
 An example of VOSA fitting is plotted in Figure~\ref{fig:fig2}.

 The masses calculated for five components in four of our systems are in the 0.06-0.07 M$_{\odot}$ range,
 what marks the substellar limit. BDs are substellar objects, that do not have enough
 mass for hydrogen burning, and burn deuterium instead. 
 With masses under 0.06-0.07 M$_{\odot}$, they can not achieve
 the temperature needed to destroy lithium 
 (Rebolo et al. 1996; Chabrier \& Baraffe 2000; Basri et al. 2000).
 Lithium is so preserved independently of the object’s age unlike VLM stars.
 But young VLM stars that still have not deleted their lithium content may
 be mistaken with BDs. By knowing the age of the object and if they
 present lithium or not, a BD can be discriminated from a young VLM object.
 Therefore, without age or lithium information, we have to consider that some of the five candidates
 may be BDs.

\subsubsection{Spectral types}

 To infer the spectral types, we used the relation between $(r-z)$ colour index and spectral type 
 given in Table~2 of West et al. (2011).
 Since 2MASS J18163409-1246310 and 2MASS J18163485-1246421
 do not have SDSS colours, we used $(J-Ks)$ relation instead, 
 and calculate the distances from the spectral type derived (see Sect 4.2.1).

  Two candidates have spectral classification from literature, 
 2MASS J13570535+3403459 and 2MASS J09332493+3232033, both classified as M6 by West et al. (2008)
 using the HAMMER stellar spectral-typing facility (Covey et al. 2007) on their SDSS spectra. 
  These values are in agreement with our photometric spectral types (M7 and M6.5, respectively).

 The spectral types from SLoWPoKES for 2MASS J01463861+1545371 / 2MASS J01463893+1545360 (M4+M4) and
 from WDS for 2MASS J13104398+1434338 / 2MASS J13104431+1434326 (M7+M8) and
 2MASS J22492429+0517137 / 2MASS J22492577+0516592 (M4+M6) are in agreement with our classification.
 The spectral type in WDS for 2MASS J18163409-1246310 / 2MASS J18163485-1246421 is given as +K:,
 while we obtained M2+M4.5.

 Spectral types are given in column 2 of Tables~\ref{tab:para} and~\ref{tab:parab}. 
 They were used to identify the primary components of the systems, defined as those with the 
 earliest spectral type.

\subsection{Properties from spectroscopy}

 To complement the photometric analysis of our candidates we got spectra for some candidates and
 searched for available spectroscopic information in public archives.

\subsubsection{Observations}

\begin{figure*}
   \centering
   \includegraphics[width=8.5cm,clip]{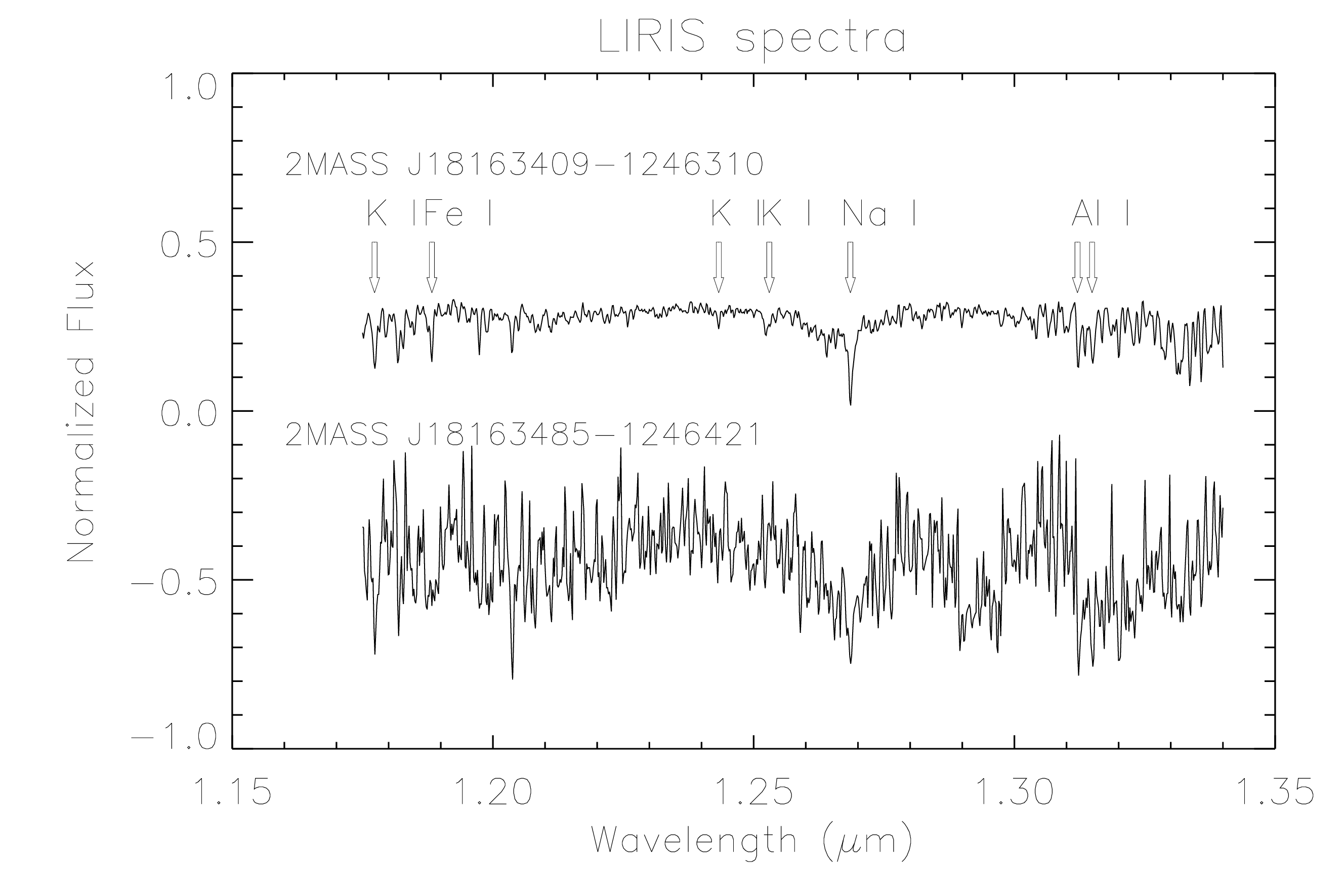}
   \includegraphics[width=8.5cm,clip]{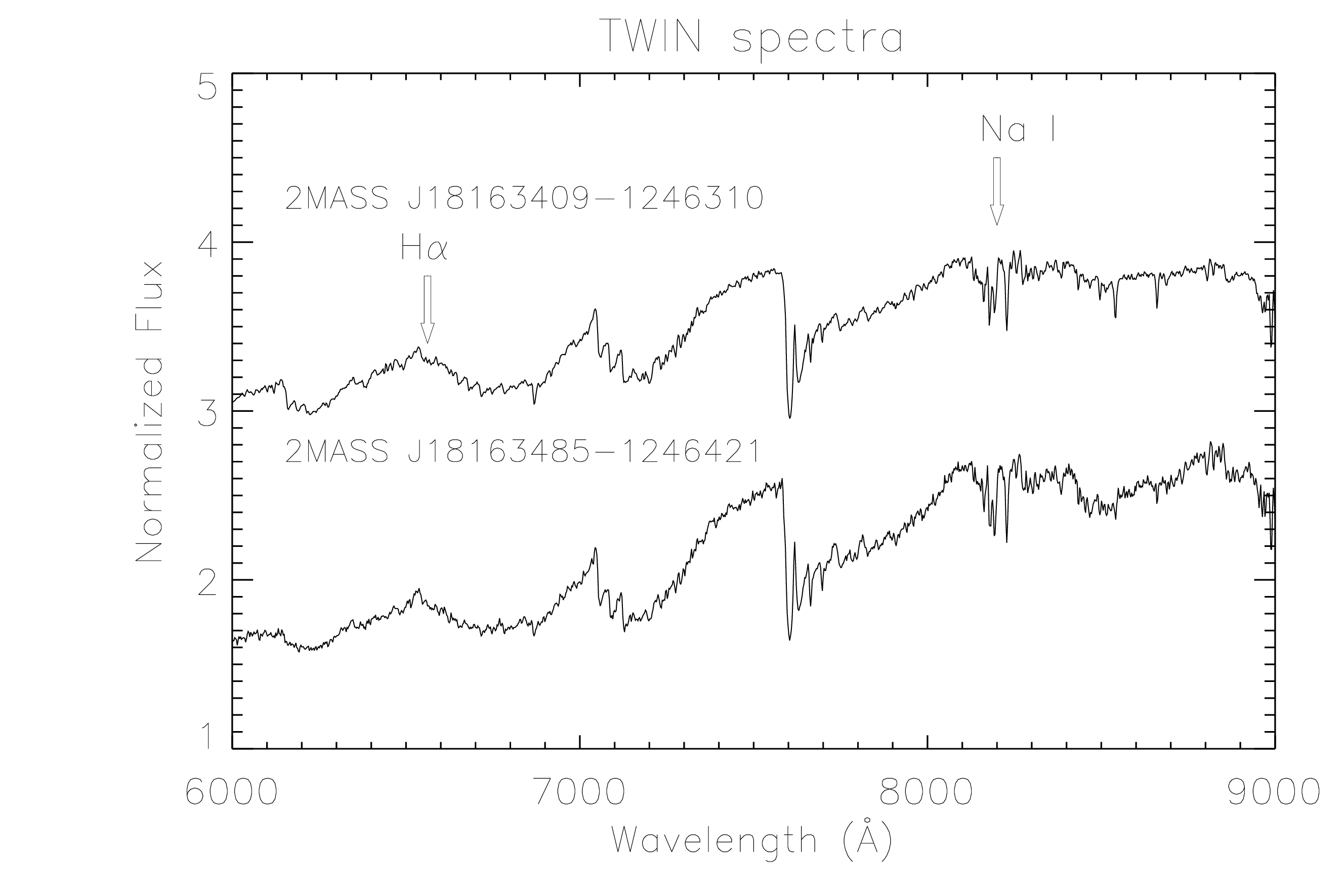}
      \caption{Observed spectra of the pair 2MASS J18163409-1246310 and 2MASS J18163485-1246421.
 Left: LIRIS IR spectra (normalized to unity and shifted for clarity).
 Right: TWIN optical spectra (normalized at 7500 \AA \ and shifted for clarity).
 Some photospheric lines (see Sect. 4.3) are marked in the figure.
}
        \label{fig:liris}
   \end{figure*}

 For the pair 2MASS J18163409-1246310 and 2MASS J18163485-1246421, we got in service mode a LIRIS
 (Long-slit Intermediate Resolution Infrared Spectrograph; Manchado et al. 1998)
 $J$ band spectra, at the 4.2~m William Herschel Telescope (WHT),
 in La Palma Observatory.
 The spectra were obtained on July 11th 2011 with a clear sky and a seeing of 1.5'',
 and were reduced in the standard way (sky subtraction,
 flat-field division, extraction of the spectra  and wavelength calibration) with
 IRAF\footnote{IRAF is distributed by the National Optical Observatory,
 which is operated by the Association of Universities for Research in
 Astronomy, Inc., under contract with the National Science Foundation.}.
 The spectra have a wavelength coverage of 1.17-1.30 $\mu$m with a
 resolving power $\lambda/(\Delta\lambda$) $\approx$2000.
 The signal to noise is $\sim$70 and $\sim$40 for each component at 1.2 $\mu$m.
 We could not make a flux calibration of the data.
 The spectra, normalized to unity, are displayed on
 left side of Figure~\ref{fig:liris}.

 We also obtained optical spectra for both components with the TWIN spectrograph
 mounted on the Calar Alto 3.5~m telescope on 26 August 2012 in service mode.
 Weather conditions were photometric and transparency was excellent with a seeing of 1 arcsec.
 The TWIN spectrograph is equipped with a 2048\,$\times$\,800 pixel CCD detector.
 We used the T11 low-resolution grating with a slit of 1 arcsec in the red arm,
 covering the 5500$-$11000 \AA \ wavelength range with a resolution of 1000 \AA.
  We installed the T13 grating in the blue arm to cover the 3500$-$5500 \AA \
 range although it offers limited use for this particular study.
 We took one exposure of 600 sec. Both components were placed on the slit.
 We reduced these optical spectra in a standard manner, using IRAF.
 We subtracted the bias and divided by the normalized
 internal flat taken just after sunset. Then, we extracted optimally
 the one-dimensional spectrum of each component in both systems. We calibrated
 our spectra in wavelength with the Helium-Argon lamps taken before sunset
 to an accuracy better than 0.1\,\AA{}.
 Finally, we calibrated the extracted spectrum of each component with a
 spectrophotometric standard (HZ44; Oke 1990) observed for this programme.
 The flux calibration is only valid up to $\sim$9000 \AA, where flux is well
 characterized. The spectra, normalized at 7500 \AA, are displayed on
 right side of Figure~\ref{fig:liris}.

\begin{figure*}
   \centering
   \includegraphics[width=8.5cm,clip]{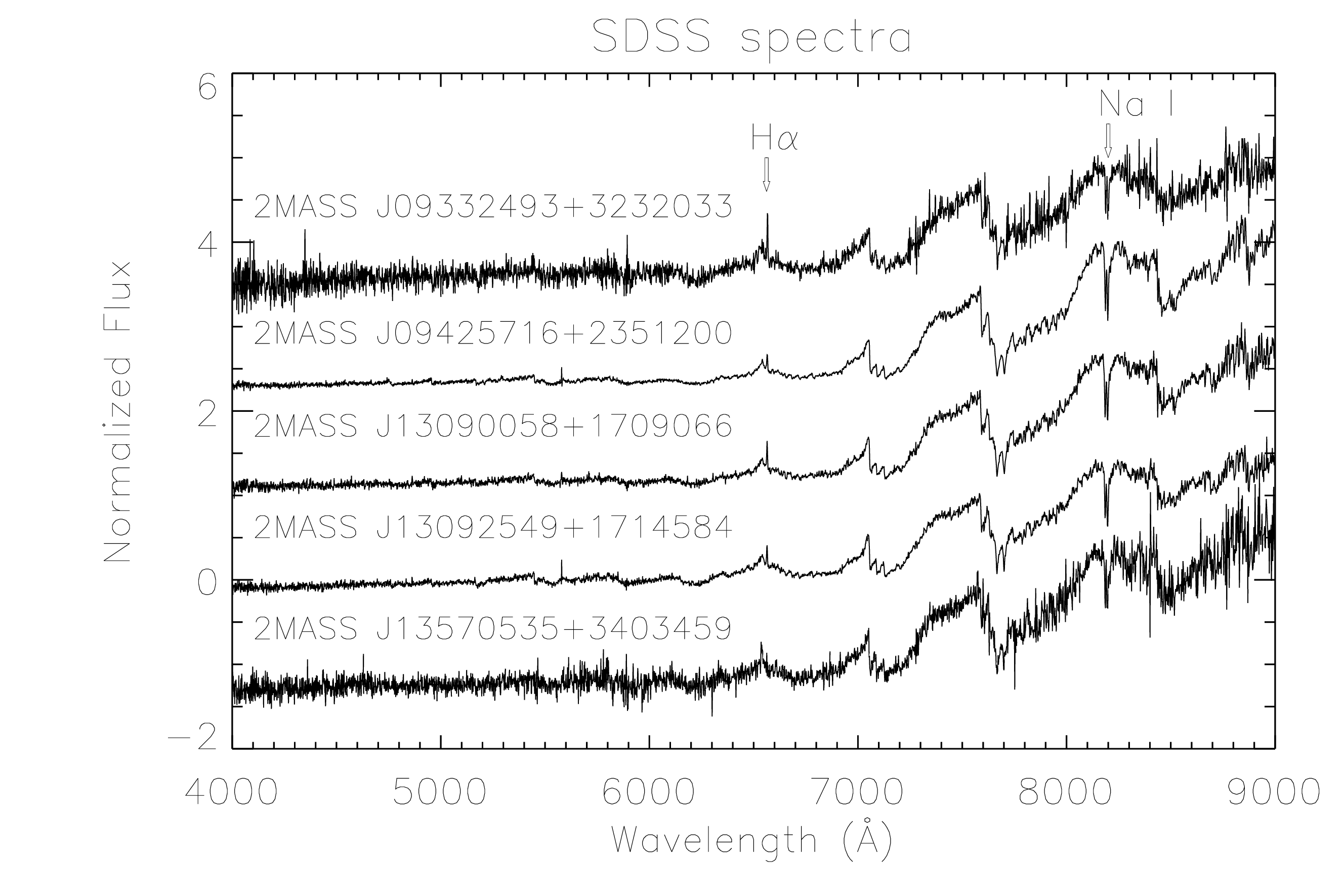}
   \includegraphics[width=8.5cm,clip]{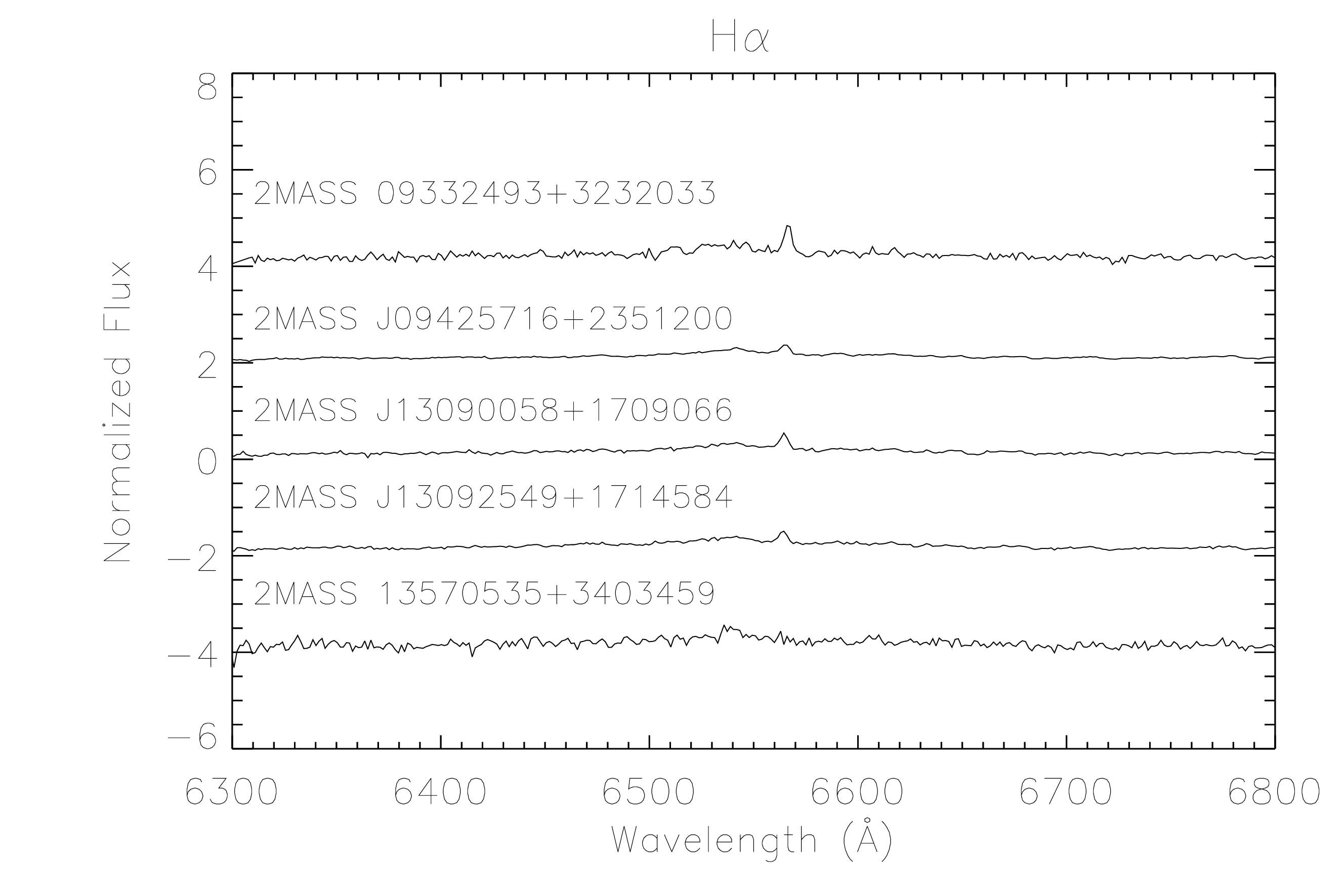}
      \caption{SDSS optical spectra (normalized at 7500 \AA \ and shifted for clarity)
 of 2MASS J09332493+3232033, 2MASS J09425716+2351200, 2MASS J13090058+1709066, 2MASS J13092549+1714584
 and 2MASS J13570535+3403459. H$\alpha$ is seen in emission in all except for 2MASS J13570535+3403459
  (zoom in the right side). Na~{\sc i} doublet ($\lambda\lambda$ 8183, 8199 \AA) is also marked.
              }
         \label{fig:espec}
   \end{figure*}

 In parallel to this, we looked for additional spectroscopic 
 information available in public archives. Only five spectra were found, all of 
 them in the SDSS data base\footnote{http://skyserver.sdss.org/dr10/en/tools/search/SQS.aspx}: 
 2MASS J09332493+3232033, 2MASS J09425716+2351200, 2MASS J13090058+1709066, 2MASS J13092549+1714584 
 and 2MASS J13570535+3403459.
 The spectra have a wavelength coverage of 3800-9200 \AA \ with a resolving
 power $\lambda/(\Delta\lambda$) $\approx$1800.
 The spectroscopic data are automatically reduced by the SDSS pipeline software.
 The spectra, normalized at 7500 \AA, are displayed in Figure~\ref{fig:espec}.

\subsubsection{Spectral types}

In the TWIN optical spectra we estimate spectral types by
 using the PC3 index as defined in
 Mart\'in et al. (1999). We obtained spectral types M4-5/M5 for
 2MASS J18163409-1246310/2MASS J18163485-1246421.
 While 2MASS J18163409-1246310 shows slightly later spectral type from PC3 than
 from photometry, 2MASS J18163485-1246421 is in good agreement  (see Table~\ref{tab:parab}).

  For LIRIS data, we could not make a flux calibration, therefore in order to get a
 spectral classification, we measured the $EW$ of K~{\sc i} doublets
 (1.168$\mu$m, 1.177$\mu$m and 1.243$\mu$m, 1254$\mu$m),
 Na~{\sc i} line (1.268$\mu$m), Al~{\sc i} doublet (1.311$\mu$m, 1.314$\mu$m) and
 Fe~{\sc i} line (1.189$\mu$m) when possible and compared them with
 Mclean et al. (2003, 2007) $EW$ diagrams and tables. 2MASS J18163485-1246421
  has a noisy spectrum, but we could measure some $EW$s. This comparison only allows us to
 determine that both components are earlier than M6 spectral type,
 which is consistent with the spectral types obtained from the TWIN spectra.

 For the SDSS data, the SDSS pipeline uses HAMMER (Covey et al. 2007) to assign spectral types
 by measuring a suite of spectral indices and performing a least-squares minimization
 of the residuals between the indices of the target and those measured from
 spectral type standards. The typical spectral type error is $\sim$ 0.5$-$1.0 subtypes.

 The spectroscopic spectral types are provided in Tables~\ref{tab:para} and \ref{tab:parab}.
  We can see how they are compatible, within the uncertainties, with the information 
 found by the photometric colours and the temperatures estimated by SED fitting.

 \subsubsection{Activity}

 Figure~\ref{fig:espec} presents the SDSS spectra (left) with a zoom in H$\alpha$
 region (right), that can be seen clearly in emission for most of them.

 We measured $EW$(H$\alpha$) in the SDSS spectra, finding values of 
 8.2 (6.63$\pm$0.0.49 by West et al. 2008), 5.5, 4.7, 4.3  \AA \ for
  2MASS J09332493+3232033, 2MASS J09425716+2351200, 
 2MASS J13090058+1709066, and 2MASS J13092549+1714584 respectively, and no emission for
 2MASS J13570535+3403459, although the line appears to be filled-up
 (West et al. 2008 give a value of $EW$(H$\alpha$)=-0.46$\pm$0.71).
 Taking into account the empirical upper limit boundary of chromospheric activity
 derived by Barrado y Navascu\'es \& Mart\'in (2003), our $EW$(H$\alpha$) 
 measurements are under the dividing line between chromospheric activity
 and disc accretion (see e.g. Valdivielso et al. 2009; Aberasturi et al. 2014). 

  Since TWIN spectra have low resolution, we could not measure
 $EW$(H$\alpha$) for 2MASS J18163409-1246310 and 2MASS J18163485-1246421.

 \subsubsection{Ages}

 Following Table~2 in West et al. (2008), M5-M7 spectral type
 objects present activity lifetimes of about 7-8 Gyr.
 Since that is an ample range, to further constrain ages of the objects
 where we could measure a value of $EW$(H$\alpha$) (2MASS J09332493+3232033,
 2MASS J09425716+2351200, 2MASS J13090058+1709066, and 2MASS J13092549+1714584),
 we compare these values and the respective spectral types with the samples in
 Shkolnik et al. (2009) and G\'alvez-Ortiz et al. (2014).
 A rough estimation associates the target activity level with an age of $\approx$ 
 100-600 Myr. In the case of 2MASS J09332493+3232033 we also have some kinematics
 from West et al. (2008) that may indicate an older age.
 The only case in which we have the spectra of both components
 with $EW$(H$\alpha$) are for the 2MASS J13090058+1709066 / 2MASS J13092549+1714584 pair.
 They show similar range age, 400-600 Myr, what may confirm they formed at the same time,
 but 2MASS J13090058+1709066 is classified as M7.5 and therefore the activity-age
 relation is in the limit of applicability.

  In the case of 2MASS J18163409-1246310 / 2MASS J18163485-1246421, the lack of H$\alpha$
 emission may indicate an old age, although Mart\'in et al. (2010) showed that for an object with a 
 spectral class between M6 and L4 its detection is not required to be classified as young.

 We also studied the Na~{\sc i} doublet ($\lambda$ 8183, 8195 \AA) for all spectra
  and K~{\sc i} doublets (1.168$\mu$m, 1.177$\mu$m and
 1.243$\mu$m, 1.254$\mu$m) for 2MASS J18163409-1246310 / 2MASS J18163485-1246421 pair.
 Schlieder et al. (2012) presented a study of the  Na~{\sc i} doublet  $EWs$ in giants, 
 old dwarfs, young dwarfs, and candidate members of the $\beta$ Pic moving group 
 using medium resolution spectra. They concluded that the Na~{\sc i} doublet can be used as 
 an age indicator for objects with spectral types later than M4 and younger than 100 Myr, where metallicity
  has an important role. Other similar studies used the sodium diagnostic to discriminate 
 objects up to 200 Myr (e.g Barrado y Navascu\'es 2006, G\'alvez-Ortiz et al. 2014).
 We measured the Na~{\sc i} pseudo-equivalent widths, $EW$=5.6/7.0/6.8/5.8/7.2/6.2/7.2 \AA,
 for 2MASS J18163409-1246310, 2MASS J18163485-1246421, 2MASS J09425716+2351200, 
 2MASS J13092549+1714584, 2MASS J13090058+1709066, 2MASS J09332493+3232033 and 2MASS J13570535+3403459,
 suggesting ages older than $\sim$200 Myrs (Mart\'in et al. 2004, see Table~2 and Figure~4 therein;
 Schlieder et al. 2012; G\'alvez-Ortiz et al. 2014, Table~10).
 Differences in  metallicity and measurements in different 
 resolution spectra, lack of homogeneity, etc, should be taken into account when
 assessing the results of these comparisons.
 In the 2MASS J18163409-1246310 / 2MASS J18163485-1246421 pair, the K~{\sc i} lines seem to
  be weaker than expected for an object with the spectral type 
 estimated using photometry. This could be an indicator of low gravity (and, therefore, 
 youngness). Nevertheless this result must be taken with caution based on the low S/N of
 the spectrum and the discrepancies found between the spectroscopic and photometric spectral types.

 Further spectral characterization where age
 estimation could be derived for each system component could state 
 if a pair has formed at the same epoch and therefore enlight the formation 
 processes of these wide low-mass systems.

\subsection{Orbital period}

 Because of projection effects, the real separation between components is expected to be,
 on average, 1.4 times larger (Couteau 1960). But in our case, we decided to
 take just the average of the measured projected separation
 (column 9 of Tables~\ref{tab:para} and~\ref{tab:parab}), and the masses
 found with VOSA (column 5 of Tables~\ref{tab:para} and~\ref{tab:parab}) to estimate an orbital
 period for each system. If VOSA mass was not available we used the mass derived from
 Kraus \& Hillenbrand (2007; column 6 of Tables~\ref{tab:para} and~\ref{tab:parab}).
 The orbital periods range from $\sim$3x10$^3$ to $\sim$5x10$^7$ years for our candidates.
 Therefore, none of our candidates will have measurable orbital motions and
  hence will have common proper motion on the sky.

\subsection{Binding Energy}

 Low mass and VLM wide binary systems are expected to have very low (absolute values of)
 gravitational potential (binding) energies, $U_{g}$ = - $GM_{1}M_{2}/r$ (with
 M$_{1}$ and M$_{2}$
 the masses of each component and r the distance between them).
  We have calculated the binding energies for our systems using the value of masses
 obtained by VOSA, and the projected physical separation (instead of the true separation)
 and compare them with those of other low and VLM, 
 wide separation binaries from the literature
 (see the last column of Tables~\ref{tab:para} and~\ref{tab:parab} and right-hand panel of
 Figure~\ref{fig:fig3}).
 We consider a physically bound system when binding energy is over
  10$^{33}$ J (Dhital et al. 2010).
  Nine of the system show energies under this limit though.
 We discuss if these candidates are bound or not in Sect.7.

\section{Newly identified L-T objects}

 In addition to the 47 very low-mass systems, we identified 50 isolated objects 
 fulfilling the L-T photometric cuts described in Sect. 2 and not 
 previously reported in the literature. We measured some characteristics
 of these objects and included them here for future studies.

 Proper motions of these targets were measured in a similar way to that 
 described in Sect. 3. but this time using only 2MASS and SDSS epochs 
  (Table~\ref{tab:obsL}, columns 7, 8).
 We used the position of the candidates in other catalogues (UKIDSS, $WISE$) 
 to confirm the modulus and sign of each component of the proper motion
 to keep them as candidates.

 Spectral types and distances were determined using the $(i-z)$ and $(i-J)$ relations 
 with spectral type and absolute magnitude given in Schmidt et al. (2010, Table~3 and eq. 1, 2). 
 Averaged values of spectral types and distances obtained with each colour are 
 given in columns 8 and 11 of Table~\ref{tab:obsL}.

 Effective temperatures and surface gravities were computed using VOSA in the same way as 
 we did for M dwarfs (Sect. 4.2.3).

 Three of the targets, 2MASS J105804+133947, 2MASS J113052+163801 and 
 2MASS 13084263+0432441 may correspond to low metallicity Ls, or L subdwarfs.
 L subdwarfs are fully convective objects representing the low mass end of 
 the Population II objects. They retain the original chemistry of the early 
 Galaxy and, therefore, are good tracers of the Galactic halo formation and evolution.
 We have marked the three targets in Table~\ref{tab:obsL} with proper motion 
 over than 150 mas yr$^{-1}$ whose $(J - K)$ and $(z-J)$ colours follow 
 the criteria of Lodieu et al. (2010) for the
 selection of this type of objects and also lay in the appropriate area of 
  the  $(i - J)$ vs $(J - K)$ diagram of Zhang et al. (2017).
 Moreover, although the reduced proper motion diagram (Figure~\ref{fig:hr}) seems 
 to not discriminate between dwarfs and subdwarfs in the L spectral type region, 
 the three targets lay close to other L subwarfs identified in the literature.

\begin{figure}
   \centering
   \includegraphics[width=8.5cm,clip]{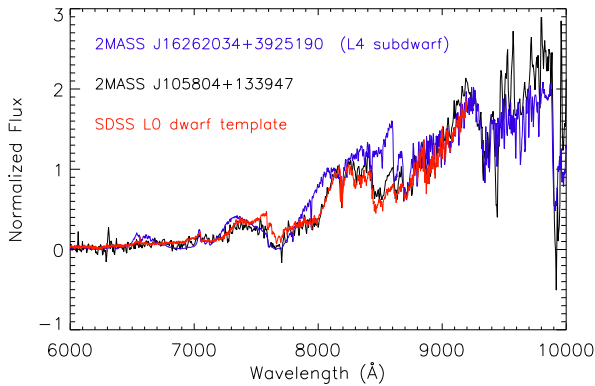}
      \caption{Comparison of 2MASS J105804+133947 spectrum with a L0 dwarf template from SDSS
 and a known subdwarf 2MASS J16262034+􏰉3925190 (Burgasser 2004).
              }
         \label{fig:especlssub}
   \end{figure}

 We observed two of these L subdwarf candidates (2MASS J105804+133947 and 2MASS J113052+163801)
 with the Optical System for Imaging and low Resolution Integrated Spectroscopy (OSIRIS; Cepa et al. 2000)
 instrument on the 10.4~m Gran Telescopio de Canarias (GTC) on 10 June 2015 in 
 service mode as part of a filler programme (programme number GTC38\_15A; PI Lodieu).
 We obtained a single on-source spectrum of 1800 sec for each candidate with the 
 grism R500R and a slit of 1 arcsec to cover the 5000--10000 \AA \ range. Bias, flat-field, 
 and arc lamps were observed during the afternoon preceding the observations as part of 
 the GTC standard calibration plan. We reduced the data under the IRAF environment 
 (Tody 1986, 1993). We removed the median-combined bias, divided by the normalized 
 median-combined dome 
 flat before extracting optimally the spectrum and calibrating it in wavelength with an rms 
 better than 0.35 \AA. We found that 2MASS J113052+163801 is probably a $\sim$M9 type dwarf whereas
  2MASS J105804+133947 looks like an early-L dwarf with some features 
 indicating metallicity less than solar. Figure~\ref{fig:especlssub} presents a 
 comparison of the optical spectrum of 2MASS J105804+133947 to SDSS templates
 of solar and low-metallicity concluding that it is most likely a dL/sdL1$\pm$1.0. 
 Further analysis is therefore needed to confirm or not its subdwarf nature.

\begin{figure}
   \centering
   \includegraphics[width=8.5cm,clip]{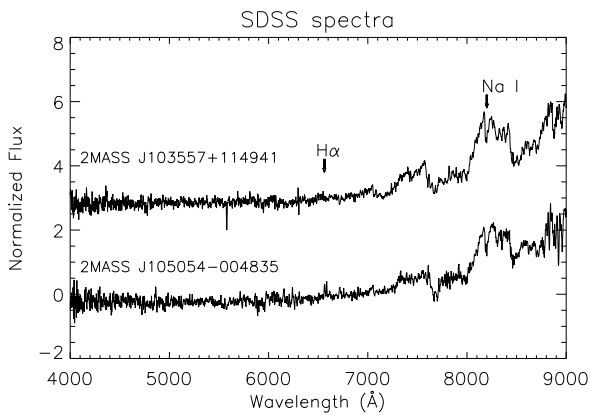}
      \caption{SDSS optical spectra, normalized at 7500 \AA \, smoothed and shifted for clarity,
 of 2MASS J103557+114941 and 2MASS J105054-004835. Both are classified as L1.5. H$\alpha$
  and Na~{\sc i} doublet region are also marked.
              }
         \label{fig:especls}
   \end{figure}

 In the same way as for M dwarfs, we searched
 for available spectroscopic information in public archives.
 We found that two of the targets, 2MASS J10355745+1149420 and 2MASS J10505470-0048352,
 have a SDSS spectrum (see Figure~\ref{fig:especls}). Both spectra are classified 
 by the SDSS pipeline software as “L dwarf”, 
 in agreement with our classification (see Table~\ref{tab:obsL}).

 
\section{Higher-mass companions}

 Very low-mass objects in multiple systems with higher-mass stars are very interesting
objects as their physical parameters can be accurately constrained via their primaries and,
therefore, can be used for testing theoretical models of atmospheres and interiors. 
 Also, low-mass wide binaries with a high-mass companion will have strongest binding
energy and, therefore, more probability to survive with time.

 We searched for higher-mass companions to the system candidates
 using Tycho-2, PPMXL and UCAC4 (Fourth U.S. Naval Observatory CCD Astrograph Catalogue;
 Zacharias et al. 2013) catalogues.
 We looked for objects whose proper motions and distances differ by less than 3$\sigma$ from the 
 average values of our pairs. Separations translating into binding energies below 10$^{33}$ J were discarded.
 For the search in PPMXL and UCAC4, we also add the condition $J$(2MASS)$<$15.5.

 With these criteria, we identified in PPMXL six objects as potential third components to 
 five systems. Searches in Tycho-2 and UCAC4 did not return new candidates.
 VOSA was utilized to obtain their temperatures.
 We considered the masses from Gray (1992) and Kaltenegger \& Traub (2009) 
 for the spectral types in the A0V-K7V and M0-M9 ranges, respectively.
 We used Table~3 of Covey et al. (2007) to derive absolute $J$-band magnitudes and
 calculate distances and separations.
 Targets and their properties are displayed in Table~\ref{tab:paratertiaries},
 marked as blue dots in Figure~\ref{fig:fig3} and discussed in next section.

 For the other binaries, we assume that, under our search conditions, 
  they are not members of higher multiplicity systems.
 Apart from the possibility of fainter or/and wider companion,
 it is also possible that some of our objects
 are unresolved binaries themselves, formed by low-mass companions of similar spectral type,
 and therefore with a total higher mass, making binding energies 
 even higher than those calculated in Section 4.5.
 
 We proceeded in a similar way and looked for possible bright companions to L-T dwarfs 
 included in Table~\ref{tab:obsL}. Only one high-mass object, PPMXL 4077732287929300487, was identified 
 as possible primary. The new system characteristics are displayed in Table~\ref{tab:paraLbinaries},
 where we assume 0.075 M$_{\odot}$ for the L dwarf in the binding energy calculations. 
 The system is marked as a green dot in Figure~\ref{fig:fig3} and discussed in next section.

\section{Discussion}

The combination of the catalogues that we used in the search allows us to be 95\% complete up to
 r=22.2, i=21.3 and z=20.5 (SDSS completeness) and J=15.8 (from 2MASS).
 The maximum distance at which an object will be included in the search is $\sim$75-700 pc for M dwarfs
 and $\sim$10-65 pc for L-T dwarfs.
 Given the angular separation limits of our search algorithm, we do expect to be overlooking genuine wide
 binaries with very large angular separations. Since we take these limits in order to keep a relative
 binding energy that guarantees real linkage of the components, we expect the percentage of binaries that
 we may miss is small.
 We have not attempted to account for incompleteness or bias, therefore the following discussion
 does not present conclusions about the population as a whole, but a comparison of our sample
 characteristics to the literature findings.

\begin{figure}
   \centering
   \includegraphics[width=8.5cm,clip]{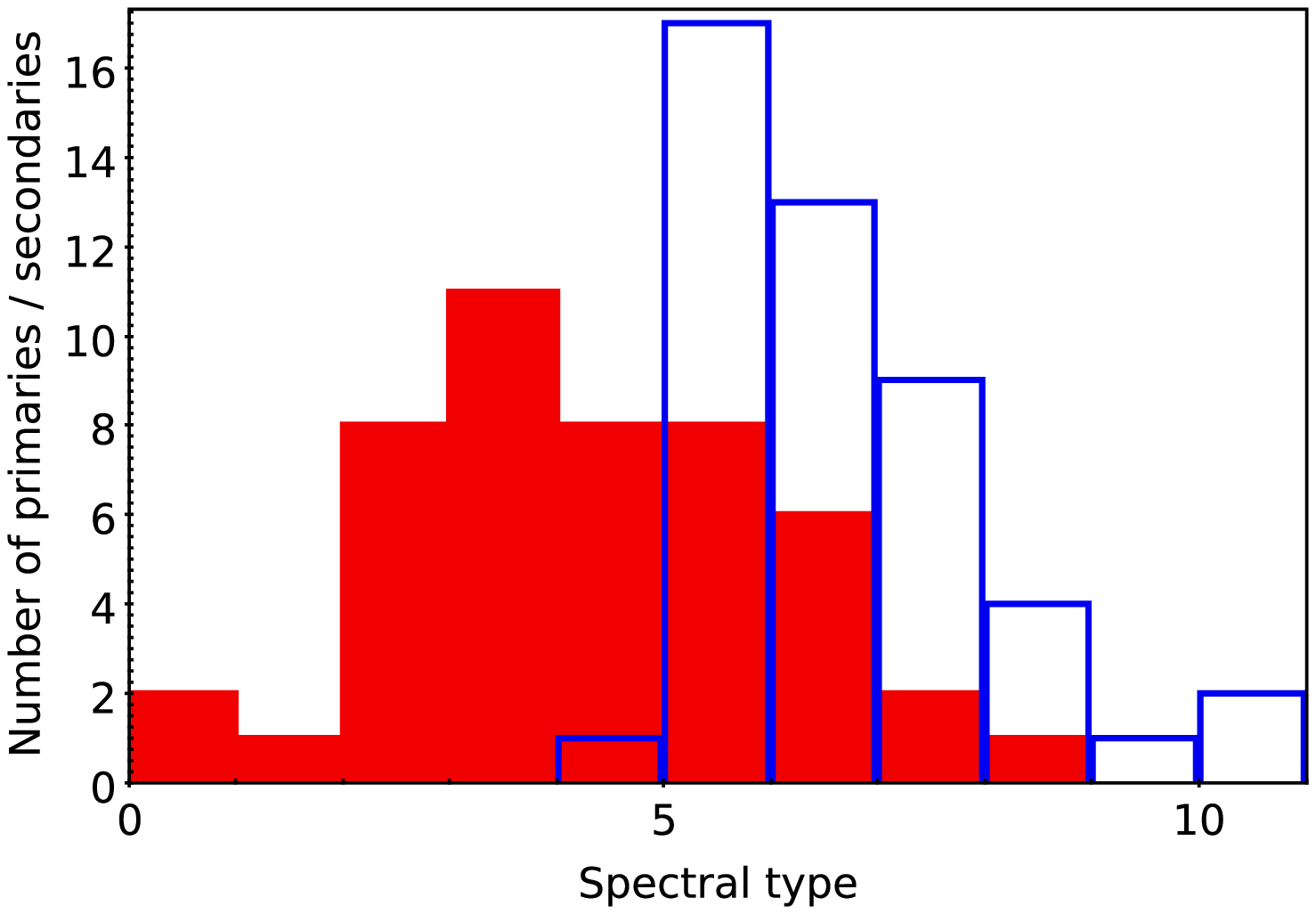}\\
   \includegraphics[width=8.5cm,clip]{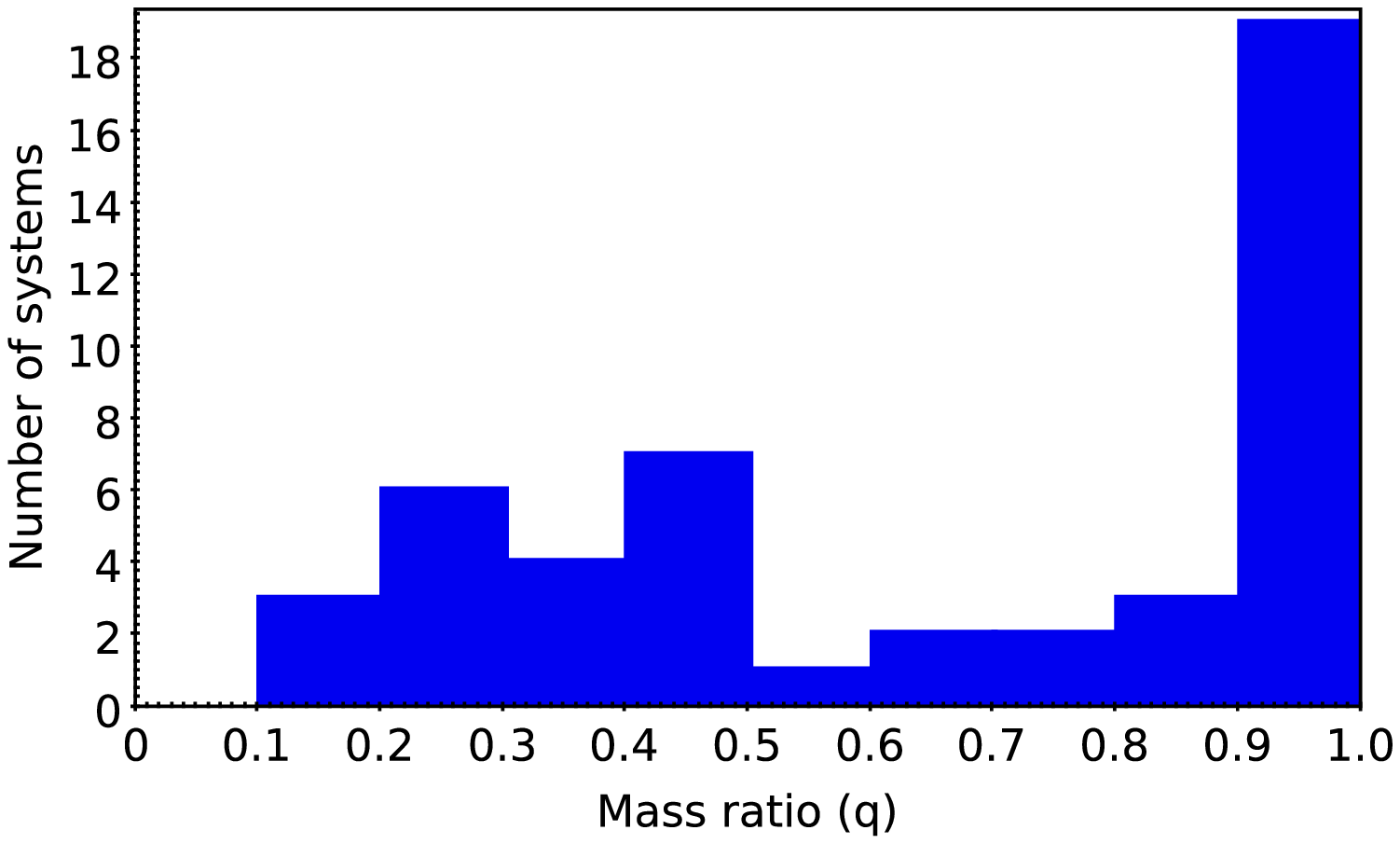}\\
   \includegraphics[width=8.5cm,clip]{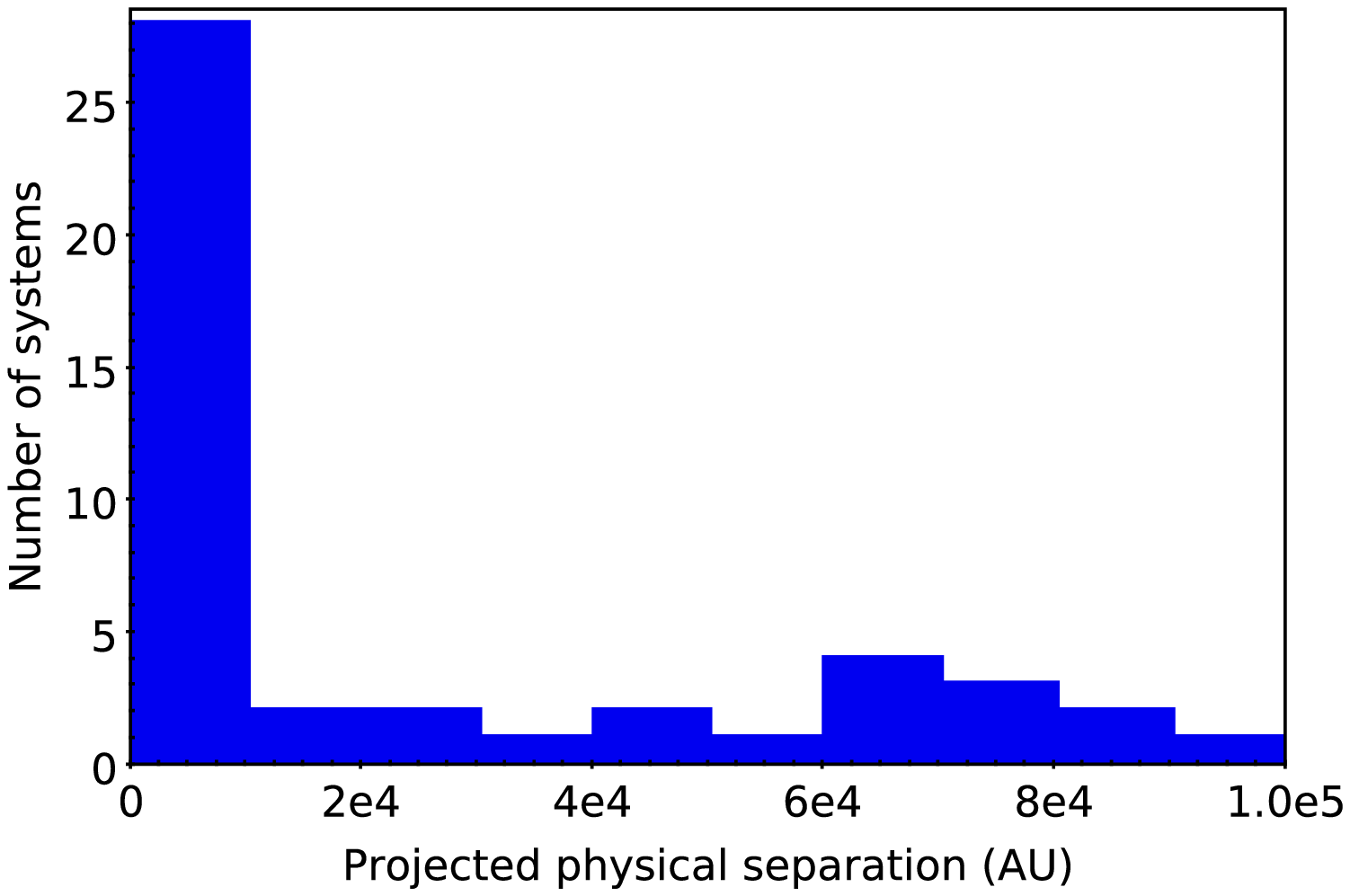}\\
      \caption{Top panel: spectral distribution of primaries (red) and secondaries (blue) 
 of the objects included in Tables~\ref{tab:obs} and ~\ref{tab:obs2}.
 Spectral types M0-L0 are represented by 0-10.
Middle panel: mass-ratio distribution for the sample. Bottom panel: projected physical separation
  distribution for the sample. The sample has not been corrected for incompleteness. 
              }
         \label{fig:hist}
   \end{figure}

 It is well known that the binary frequency decreases with primary mass, i.e. spectral type
 (see e.g. Aberasturi et al. 2014; Figure~9).
 Therefore the distribution of primaries (defined as the component with the earliest 
 spectral type), should show a smooth decrement towards later spectral types with
 a peak around M3-M4 (e.g. Farihi et al. 2005; Reid et al. 2007; Stelzer et al. 2013, and references therein). 
 We plot in Figure~\ref{fig:hist} (top panel), the distribution of primaries (red) and secondaries (blue)
 for our sample. They follow the expected trend.

The mass-ratio and semimajor axis distributions for binary components provide key information
 about the formation mechanism of the systems. The mass ratios of binaries among Sun-like
 stars are distributed quite uniformly, while for BDs a preferentially nearly equal
 in mass distribution is present (e.g., Burgasser et al. 2007; Raghavan et al. 2010; Bate 2012).
 For our small sample, we find a rather uniform distribution (Figure~\ref{fig:hist} middle panel). 
 The peak in the 0.9-1.0 interval may reflect the presence of VLM/BD binary systems among our candidates.
 Also, as expected from VLM systems, the projected physical separation distribution peaks at small values 
  ($<$ 10$^{4}$ AU, Figure~\ref{fig:hist} bottom panel), with a maximum between 500-1000 AU. 
  The shape of this histogram, showing an increasing trend after $\sim$ 30000 AU, may indicate
  chance alignment contamination at larger separations.
  In any case the evolution of the multiplicity parameters with mass and its impact on the formation 
 mechanisms of stars and BDs requires a statistically significant and unbiased sample and 
 it is beyond the scope of this paper.

\begin{figure*}
   \centering
   \includegraphics[width=8.5cm,clip]{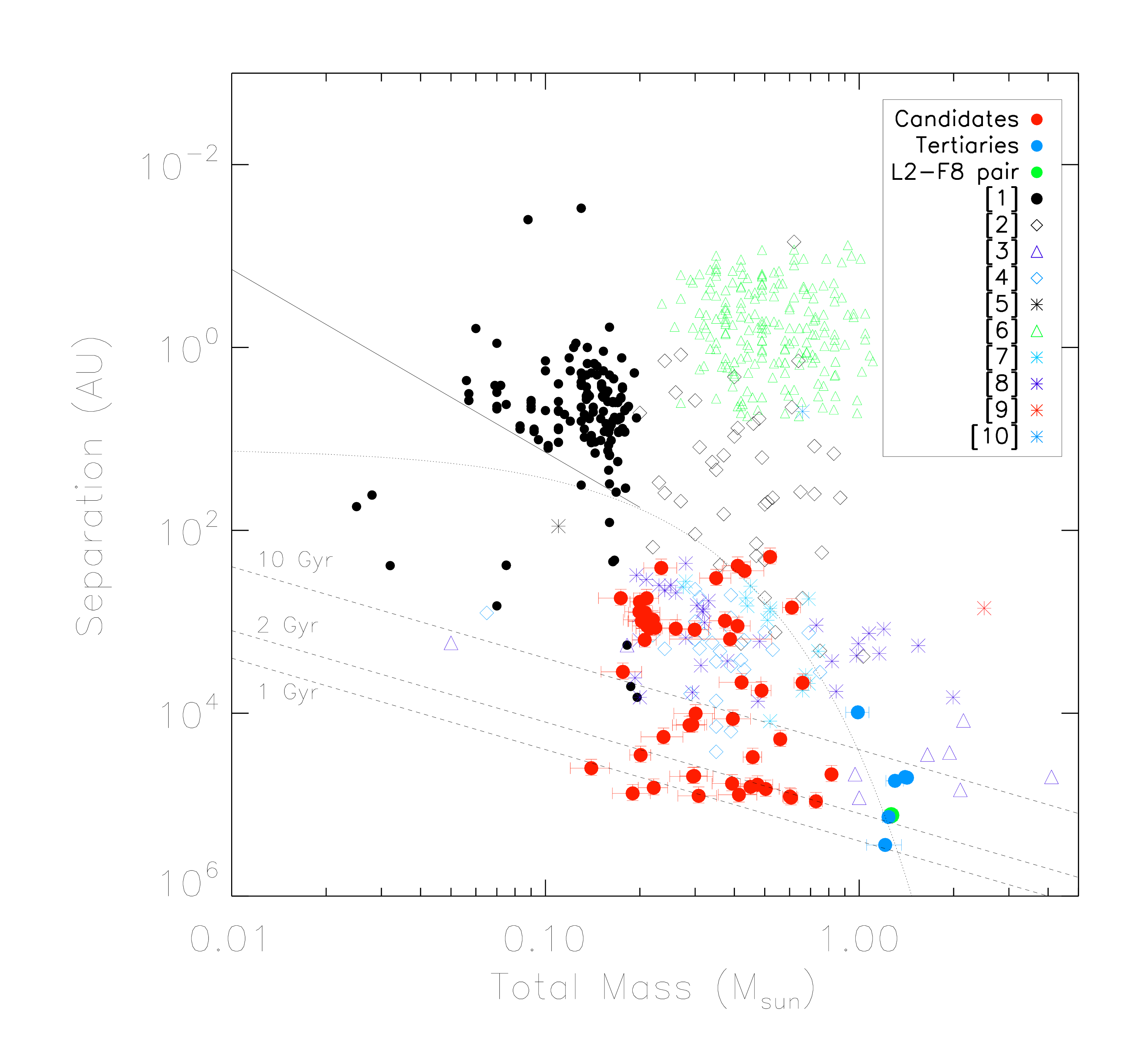}
    \includegraphics[width=8.5cm,clip]{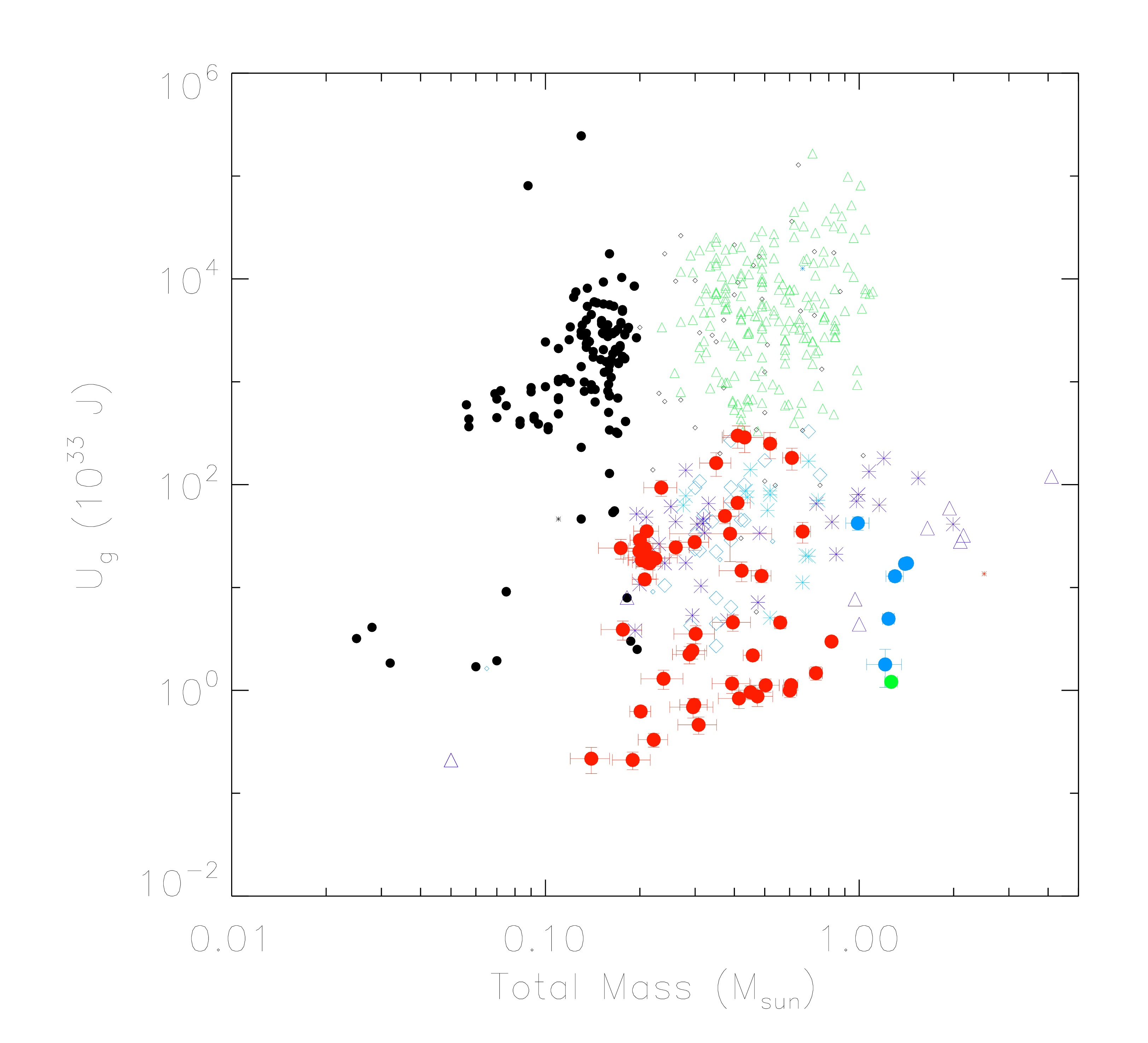}
      \caption{Separation and binding energy vs. total system mass.
 Known binary systems from the literature are marked with different symbols and
 our candidates to M-M, M-M with a higher mass companion and the only pair found for
 an L dwarf, are plotted with filled circles in red, blue and green respectively.
 In the left-hand panel we also overplot
 the empirical limit for stability stated by Reid et al. (2001) (dotted line) and
 by Burgasser et al. (2003) (solid line). We also overplot the
 lifetime “isochrones” suggested by Dhital et al. (2010) with data of Close et al. (2007) and
 Weinberg et al. (1987) equations for dissipation times of 1, 2, and 10 Gyr (dashed lines).
  Figure legend: [1]: VML archive, Burgasser et al. (2007) and Siegler et al. (2005), 
 [2]: Fischer \& Marcy (1992) and Reid \& Gizis (1997),
 [3]: Caballero et al. (2006) \& Caballero (2007, 2009), 
 [4]: Luhman et al. (2009, 2012), Dhital et al. (2010) \& Mu$\breve{z}$i\'c et al. (2012),
 [5]: Burningham et al. (2010), [6]: Janson et al. (2012),
 [7]: Deacon et al. (2014), [8]: Faherty et al. (2010) \& Baron et al. (2015),
 [9]: HIP 78530 AB, Lafreni\`ere et al. (2011),[10]: Ross 458 AB, Goldman et al. (2010).
   }
         \label{fig:fig3}
   \end{figure*}

  We plot in left-hand panel of Figure~\ref{fig:fig3}, a mass-separation diagram
 of our candidates, including previously known VLM binaries and some
 higher mass binaries, while in the right-hand panel,
 binding energy versus mass is plotted.
 From the 47 systems, 15 present separations in the 100-1000 AU range, 13 in 1000-10000 AU 
 and 22 over 10000 AU. There are nine pairs with binding energies under the 10$^{33}$ J limit.

 {\it  Are these systems bound? how could they survive disruption?
 for how long are they going to be bound?}
 According to Close et al. (1990) "the physical limit to the maximum 
 separation [that] a candidate system could possibly have is the separation at 
 which the differential Galactic force exceeds the gravitational binding force of the system".
 To study the systems stability, we used and compared different literature sources.

 Following the Weinberg et al. (1987) work, Reid et al. (2001) and Burgasser et al. (2003)
 quantified different empirical relations between the maximum separation limit for
 VLM binaries in the field and their total mass. 
 The lines representing these limits are plotted also in Figure~\ref{fig:fig3} left-hand panel.
 As can be seen in the figure, Burgasser et al. (2003) is only applicable to a small
 number of targets in our sample (The relation is only valid for 
 systems with M$_{\textit{tot}}$ $<$ 0.2 M$_{\odot}$), none in the stability area, 
 and only four are inside the stability limits marked by Reid et al. (2001).
 We should remark that Reid and Burgasser relations are empirical and based in samples with
 none or few such wide binaries, a number that increased after these works, 
 implying maybe that they are out of date and probably too restrictive. 

 More recently, Dhital et al. (2010), applied Close et al. (2007) Galactic disc mass density
 and other updated parameters to Weinberg et al. (1987) equations.
 They applied the resultant equation (Dhital et al. 2010; eq.18) to
 describe statistically the widest binary that is surviving at a given age. 
 In Figure~\ref{fig:fig3} letf panel, we overplot the lifetime ''isochrones'' suggested 
 by Dhital et al. (2010) for dissipation times of 1, 2, and 10 Gyr. 
 Twenty seven of the 51 pairs of the sample are situated over the 10 Gyr limit,
 eight are situated between the 2 and 10 Gyr isochrones, 12 between the 1 and 2 Gyr 
 isochrones and four pairs are under the 1 Gyr isochrone limit.

{\it Do these results mean that the pairs that do not accomplish the stability criteria 
 are in the process of disruption or already separated?}
 The answer is complicated, since for example, the age of the system would be needed. 
 Dhital et al. (2010) find a {\it bimodality} in the separation of VLM binaries.
 They argued that the finding was real and not due to any contamination
 or bias in the sample, and it is also predicted in recent N-body simulations
 (e.g. Jiang \& Tremaine 2009; Kouwenhoven et al. 2010). They
 suggested that this bimodality reveals two distinct populations
 of wide binaries, possibly representing systems that form and/or
 dissipate through differing mechanisms (see references therein for details).
 Following this suggestion, we find also the possible division of our sample in two population:
 one tightly bound wide systems that are expected to last more than 10 Gyr,
 and other formed by weak bound wide systems that will dissipate within a few Gyrs.

 An estimation of the system ages is, therefore, needed to finally assess if they
  are actually bound or already disrupted.



 Other issue to discuss is the minimum binding energy that should be expected.
 We established a limit of 10$^{33}$ J in our initial conditions to consider
 a common proper motion pair as candidate. But this restriction has been based 
 in the typical minimum values obtained from other systems in the literature like Dhital 
 et al. (2010). Final results give nine pairs slightly under this limit, 0.2-0.9 x 10$^{33}$ J.
 We decided to keep them since the error in masses could easily situate
 them in higher energies and take into account the existence of objects such as
 SE 70 + S Ori 68 pair found by Caballero et al. (2006) with 0.2 x 10$^{33}$ J binding energy.


 All candidate systems fulfill the same conditions of distance and proper motion
 (and low probability of chance alignment). Without accurate age determination, 
 we can not state if some of them are already under disruption processes.
 Therefore, although contamination is expected for the widest systems in our sample,
 we keep them in the list of potential wide multiple systems because our
 main goal is to explore the most extreme regions.
 

Five systems may have a high-mass companion, with spectral types $\approx$G7-K9.
 One of these systems may be a quadruple formed by two M dwarfs and two K stars,
 2MASS J09335192+3237270 (M2.0), 2MASS J09332493+3232033 (M6.5), PPMXL 4233992647964279227 (K6) and
 PPMXL 4234014784472339948 (TYC 2497-1059-1; K0). For 2MASS J09332493+3232033 we estimated an age of
  $<$ 600 Myr, through $EW$(H$\alpha$) measurement,
 although W component of Galactic velocity (West et al. 2008)
 may indicate an older age (Sect. 4.3.4). Therefore, if the objects are gravitationally
 linked as seems for the binding energy values, the rest of the components would have the same age.
 In the left-hand panel of Figure~\ref{fig:fig3},
 the M-M system lay over the 2 Gyr isochrone while their pairing with the K components are 
 between the 2 and 10 Gyr lifetime isochrones for dissipation times which indicates that,  
 according to our age estimation, the system is still physically bound.

 Also, in the system formed by the 2MASS J02014517+1124244 (M2.5), 2MASS J02020000+1115202 (M5.0) pair and
 the PPMXL 2102675187759530038 (G7), both the M-M pair and their pairing with the G object
 are laying on the 1 Gyr lifetime isochrone in Figure~\ref{fig:fig3}.
 Since we have no information about the age of the system, we can not confirm if their components
 are still bound or not.

 The other pairs with tertiaries are M-M-K triples.
 2MASS J11243764+1137085 / 2MASS J11242609+1139504 / PPMXL 4137926243272078734 masses and separation
 provide with survival times between 2 and 10 Gyrs. 2MASS J13475983+3343241 / 2MASS J13480028+3341587
 lays over the 10 Gyr isochrone while its pairing with the PPMXL 4551655942546822025 is well
 over the 10 Gyr limit. 2MASS  J13572013+0550251 / 2MASS J13565606+0552499 lays over the 2 Gyr isochrone
 and its pairing with PPMXL 4398878071445995714 is situated between the 2 and 10 Gyrs.   
 These three systems are therefore probably still bound.
 
 We also have the case of the L2.0 dwarf 2MASS J09320299+1231027 and the F8 PPMXL 4077732287929300487
 pair. With a separation of $\sim$130450 AU and a binding energy of 1.2 x 10$^{33}$ J, it lays
 over the 2 Gyr lifetime isochrone in Figure~\ref{fig:fig3}. We have no information about
 the possible age of the targets so it might already be a dissipated pair.

\section{Summary and conclusions}

 Taking advantage of VO tools, we could identify 36 new low and VLM systems
  (M0.5-L0 spectral types) with separations in the $\approx$ 200 - 92000 AU range.
  We also provided distances and masses of eleven pairs previously marked
  as binaries in the literature, giving spectral classification for seven of them 
 for the first time and improving the classification for one of them using photometry and spectroscopy.

In the mass-separation parameter space covered by our objects (separations over 
 200 AU and total system masses under 1 M$_{\odot}$) we have increased by 34\%
 the number of systems previously known (Tables~\ref{tab:wlmbfl} and~\ref{tab:slow}). 
 The systems recently found by Dhital et al. (2015) were not taken into account 
 as they require further confirmation.

 From the total 47 systems, we have seven pairs formed by two 
 low-mass objects, nine pairs formed by two VLM objects,
 29 pairs formed by one low-mass and one VLM object and for the two triples,
 one is formed by three VLM objects and the other by one low-mass and two VLM objects.  
 Common proper motions and distances, with negligible probability of
 change alignment, agree with the candidates forming a bounded pair.

 Spectral types of each component were calculated using several methods.
 They cover the M0-L0 range, with $\approx$35 of the total 95 components in M6-M8 range.
 This corresponds to the transition between low-mass stars and what is called
 ultracool dwarfs (UCDs), characterized by a dramatic change in the spectra due to the
 onset of dusty condensation (Jones \& Tsuji 1998). 

 The 2MASS J13090058+1709066 / 2MASS JJ13092549+1714584 pair masses are 0.07 M$_{\odot}$
 for both components. The spectra analysis in Sect. 4.3.4. allowed to estimate an age
 in the 400-600 Myr range, what would imply a possible BD nature of both 
 components of the system. High enough resolution spectra to determine the presence
 or not of lithium in their atmospheres would help to resolve the problem  of its stellar 
 or substellar nature.
 A confirmed binary composed by two BDs at wide separation would provide key 
 information and serve as test-bed for their formation and evolution theories. 
 There are three other objects with masses in the 0.06-0.07 M$_{\odot}$ 
 range for which, without age or lithium information, is not possible to conclude about their nature.

 Distances were estimated by using different relations in the literature showing
 a good agreement between components of each system.
 We calculated masses with theoretical isochrones and 
 evolutionary models. 

 Although previously reported as possible binary in WDS catalogue, we could analysed in more detail
 the photometric properties of the 2MASS J18163409-1246310 / 2MASS J18163485-1246421 pair. 
 Through spectroscopy we were able to obtain a reliable spectral classification for both
 components and measured spectral characteristics that helped to constrain the age of this pair.

 We also identified 50 objects that we can photometrically classify as $\approx$L0-T2 spectral
 type and that do not present low-mass proper motion companions according to our requirements.
  Three of them show proper motion and photometric characteristics 
 compatible with being low metal L subdwarfs. Two of them were spectroscopically followed-up.
  One was dismissed and the other one was tentatively classified as a 
 dL/sdL1±1.0, with further analysis needed to confirm its subdwarf nature.
 This is a very interesting result given the scarce number (36) of these objects known so far. 

 A high-mass companion search gave us six potential companions to five of the M-M pairs 
 and one to an L dwarf. These systems are excellent benchmarks to determine M dwarf characteristics.
 Many authors are actually searching for binaries consisting of an F-, G- or K-type 
 dwarfs and an M dwarf, since high-mass companion can provide information of physical parameters 
 (e.g. abundances) that are much more difficult and inaccurate to measure on the secondary.
  In future work we will follow this research line
 to further characterize the components of the systems with high-mass companions.

 The percentage of higher-order multiple system in the sample is rather small.
 We only found two multiple systems formed only by low and very low-mass components.
 One with components with M6.0, M7.5 and M8.5 spectral types, and other with components
 with spectral types M0.5, M7.0 and M8.5. With high-mass component we
 found five possible multiple systems, formed all of them by two Ms and a tertiary
 of F, K spectral types. One of them could also be a quadruple, formed by two Ms and two Ks. 

 Although the sample is small, we find that it follows a rather uniform distribution in
  mass-ratio, with and average value of 0.6 and a peak in the 0.9-1.0 interval.
 The semimajor axis distribution presents a peak at $>$ 10$^{4}$ AU (with a maximum between 500-1000 AU). 

 We find that half of the sample may correspond to tightly bound wide systems 
 that are expected to last more than 10 Gyr, and the other half to systems
 formed by weakly bound wide systems that will dissipate within a few Gyrs.
 Although we could not constrain the age of most of the candidates, probably
 most of them are still bound except four (2MASS J02014517+1124244 / 2MASS J02020000+1115202,
 2MASS J09102559+0648533 / 2MASS J09105025+0648111, 2MASS J16183619+3005336 / 2MASS J16184068+2958293
 and 2MASS J13092549+1714584 / 2MASS J13090058+1709066) formed by low-mass components whose
 lifetimes are $\sim$ 1 Gyr and so maybe already under disruption processes.

 The discovery of these low-mass, wide (up to 92000 AU) binary systems is a strong 
 evidence that such systems are not as rare as was thought.
 The parameter space analysed here has been poorly explored before with the exception
 of very few papers (see references quoted throughout in this work). 
  Gathering an statistically significant number of wide, low and VLM systems
 is fundamental in order to better understand their formation and evolution processes, 
 updating and constraining the actual theories. New, more accurate and deeper surveys are 
 necessary to provide a realistic estimation of the population of this type of objects.
 Dhital et al. (2015) catalogue can provide with a good list, and soon we will have 
  the detailed astrometrical and 3D kinematical picture of the 
 Galaxy made by Gaia\footnote{http://www.esa.int/Our$\_$Activities/Space$\_$Science/Gaia},
 that will firmly confirm or refute the true binary nature of our candidate systems.

 The VO has proved to be an excellent research methodology. 
 In particular, it allowed an efficient management of the queries to the 
 different catalogues and archives used in this work as well as to determine 
 physical parameters ($T_{eff}$ and $log~g$) through VO-tools like VOSA.

\begin{table*}
\caption[]{Candidate position data for group A. 
 2MASS, SDSS and UKIDSS.
}
\label{tab:obs}
\begin{flushleft}
\begin{center}
\scriptsize
\begin{tabular}{lcccccccc}
\noalign{\smallskip}
\hline
\noalign{\smallskip}
  \multicolumn{1}{|c|}{2MASS} &  \multicolumn{1}{c|}{RA\_2MASS} & \multicolumn{1}{c|}{DEC\_2MASS} &  
  \multicolumn{1}{c|}{RA\_SDSS} &  \multicolumn{1}{c|}{DEC\_SDSS} &
  \multicolumn{1}{c|}{RA\_UKIDSS} &  \multicolumn{1}{c|}{DEC\_UKIDSS} &
  \multicolumn{1}{c|}{[pmRA,pmDEC]} &  
  \multicolumn{1}{c|}{pm} \\
& (deg) & (deg) & (deg) & (deg) & (deg) & (deg) & (mas yr$^{-1}$) & (mas yr$^{-1}$)\\

\noalign{\smallskip}
\hline
\noalign{\smallskip}
  00384499+0519003 & 9.687477 & 5.316770 & 9.687615 & 5.316649 & 9.687654 & 5.316653 & [81 (1), -66 (2)] & 105 \\ 
  00391056+0525351 & 9.794026 & 5.426419 & 9.794192 & 5.426348 & 9.794221 & 5.426313 & [90 (2), -49 (2)] & 103 \\ 
\hline
  00421358+0431185 & 10.556616 & 4.521807 & 10.556453 & 4.521728 & 10.556358 & 4.521738 & [-87 (23), -29 (7)] & 93 \\ 
  00421418+0431200 & 10.559096 & 4.522232 & 10.558930 & 4.522152 & 10.558835 & 4.522159 & [-89 (23), -31 (6)] & 94 \\
\hline
  00513590+0735192$^{*}$ & 12.899596 & 7.588668 & 12.899483 & 7.588589 & 12.899426 & 7.588535 & [-43 (4), -55 (0.4)] & 70 \\
  00513688+0735177 & 12.903701 & 7.588265 & 12.903593 & 7.588175 & 12.903535 & 7.588117 & [-37 (5), -57 (0.1)] & 68 \\
\hline
  01463861+1545371$^{+}$ & 26.660903 & 15.760323 & 26.661022 & 15.760063 & 26.661050 & 15.760038 & 38 (1), -87 (7)] & 95 \\
  01463893+1545360 & 26.662245 & 15.760003 & 26.662362 & 15.759737 & 26.662384 & 15.759711 & [37 (0.3), -89 (7)] & 97 \\  
\hline
  01554912+0527334 & 28.954707 & 5.459282 & 28.954912 & 5.458952 & 28.954993 & 5.458881 & [65 (7), -102 (13)] & 121 \\ 
  01560037+0528494 & 29.001552 & 5.480414 & 29.001712 & 5.480140 & 29.001782 & 5.480101 & [68 (8), -114 (8)] & 133 \\ 
 & & & &  & & & &  \\
  01554912+0527334 & 28.954707 & 5.459282 & 28.954912 & 5.458952 & 28.954993 & 5.458881 & [65 (7), -102 (13)] & 121 \\
  01560053+0528562 & 29.002213 & 5.482295 & 29.002381 & 5.481999 & 29.002438 & 5.481966 & [74 (2), -137 (2)] & 156 \\
& & &  & & & & & \\
  01560037+0528494 & 29.001552 & 5.480414 & 29.001712 & 5.480140 & 29.001782 & 5.480101 & [68 (8), -114 (8)] & 133 \\
  01560053+0528562 & 29.002213 & 5.482295 & 29.002381 & 5.481999 & 29.002438 & 5.481966 & [74 (2), -137 (2)] & 156 \\
\hline
  01575409+0923371$^{*}$ & 29.475390 & 9.393665 & 29.475688 & 9.393642 & 29.475735 & 9.393582 & [89 (2), -24 (16)] & 92 \\ 
  01575468+0923422 & 29.477859 & 9.395068 & 29.478163 & 9.395026 & 29.478210 & 9.394972 & [96 (2), -30 (13)] & 101 \\
\hline
  02014517+1124244 & 30.438228 & 11.406787 & 30.438407 & 11.406635 & 30.438435 & 11.406639 & [74 (1), -52 (1)] & 91  \\
  02020000+1115202 & 30.500025 & 11.255633 & 30.500192 & 11.255512 & 30.500249 & 11.255519 & [86 (3), -42 (1)] & 96 \\ 
\hline
  03165635+0617027$^{*}$ & 49.234794 & 6.284084 & 49.235055 & 6.283981 & 49.235256 & 6.283850 & [119 (2), -128 (5)] & 175 \\ 
  03165886+0618086 & 49.245277 & 6.302416 & 49.245533 & 6.302317 & 49.245740 & 6.302179 & [118 (0.3), -115 (2)] & 164 \\
\hline
  08255151+2925220$^{*}$ & 126.464651 & 29.422785 & 126.464625 & 29.422749 & 126.464542 & 29.422597 & [-25 (1), -87 (2)] & 91 \\
  08255269+2925169 & 126.469569 & 29.421366 & 126.469563 & 29.421331 & 126.469484 & 29.421178 & [-25 (4), -85 (2)] & 89 \\
\hline
  09045044-0158572 & 136.210193 & -1.982583 & 136.210118 & -1.982526 & 136.209978 & -1.982465 & [-70 (4), 76 (5)] & 103 \\
  09045115-0159297 & 136.213158 & -1.991594 & 136.213105 & -1.991520 & 136.212964 & -1.991452 & [-73 (5), 69 (15)] & 100 \\
\hline
  09102559+0648533 & 137.606655 & 6.814812 & 137.606631 & 6.814739 & 137.606591 & 6.814623 & [-24 (1), -105 (7)] & 107 \\
  09105025+0648111 & 137.709387 & 6.803089 & 137.709378 & 6.802990 & 137.709324 & 6.802893 & [-26 (8), -109 (9)] & 113 \\
\hline
  09332493+3232033 & 143.353891 & 32.534275 & 143.353733 & 32.534163 & 143.353651 & 32.534156 & [-67 (5), -41 (16)] & 79 \\
  09335192+3237270 & 143.466368 & 32.624184 & 143.466186 & 32.624113 & 143.466069 & 32.624054 & [-83 (1), -44 (0.5)] & 94 \\ 
\hline
  09442260+0842235 & 146.094169 & 8.706537 & 146.094171 & 8.706479 & 146.094379 & 8.706393 & [119 (2), -86 (2)] & 147  \\
  09442313+0842231 & 146.096405 & 8.706420 & 146.096413 & 8.706367 & 146.096618 & 8.706278 & [125 (3), -85 (1)] & 152  \\
\hline
  10132882+1136041 & 153.370110 & 11.601140 & 153.370019 & 11.601044 & 153.369914 & 11.600973 & [-85 (3), -69 (5)] & 109 \\
  10132950+1135558 & 153.372924 & 11.598843 & 153.372840 & 11.598747 & 153.372734 & 11.598678 & [-81 (3), -65 (4)] & 104 \\
\hline
  10402749-0248092 & 160.114557 & -2.802568 & 160.114504 & -2.802573 & 160.114337 & -2.802749 & [-60 (9), -56 (3)] & 82  \\
  10402763-0248063 & 160.115158 & -2.801773 & 160.115128 & -2.801785 & 160.114953 & -2.801964 & [-59 (3), -58 (1)] & 83  \\
\hline
  11002544+1445240 & 165.106036 & 14.756691 & 165.105892 & 14.756624 & 165.105814 & 14.756529 & [-64 (2), -85 (5)] & 105  \\
  11002578+1445299 & 165.107429 & 14.758311 & 165.107289 & 14.758234 & 165.107217 & 14.758137 & [-67 (2), -76 (3)] & 101  \\
\hline
  11242609+1139504 & 171.108715 & 11.664010 & 171.108723 & 11.663971 & 171.108969 & 11.663845 & [82 (4), -77 (2)] & 112 \\
  11243764+1137085 & 171.156848 & 11.619028 & 171.156953 & 11.618929 & 171.157071 & 11.618884 & [96 (30), -63 (6)] & 114 \\
\hline
  12043718+1505286 & 181.154927 & 15.091279 & 181.154804 & 15.091178 & 181.154646 & 15.091013 & [-66 (4), -75 (1)] & 100  \\
  12050629+1508193 & 181.276223 & 15.138718 & 181.276089 & 15.138597 & 181.275988 & 15.138506 & [-94 (0.5), -86 (1)] & 128  \\
\hline
  12432885-0105208 & 190.870227 & -1.089120 & 190.869961 & -1.089238 & 190.869881 & -1.089241 & [-99 (8), -37 (6)] & 106 \\
  12433001-0105152 & 190.875080 & -1.087576 & 190.874845 & -1.087703 & 190.874763 & -1.087705 & [-107 (2), -31 (3)] & 111 \\
\noalign{\smallskip}
\hline
\noalign{\smallskip}
\end{tabular}
\end{center}
\end{flushleft}
\end{table*}

\addtocounter{table}{-1}
\begin{table*}
\caption[]{Cont.
}
\label{tab:obs}
\begin{flushleft}
\begin{center}
\scriptsize
\begin{tabular}{lcccccccc}
\noalign{\smallskip}
\hline
\noalign{\smallskip}
  \multicolumn{1}{|c|}{2MASS} &  \multicolumn{1}{c|}{RA\_2MASS} & \multicolumn{1}{c|}{DEC\_2MASS} &
  \multicolumn{1}{c|}{RA\_SDSS} &  \multicolumn{1}{c|}{DEC\_SDSS} &
  \multicolumn{1}{c|}{RA\_UKIDSS} &  \multicolumn{1}{c|}{DEC\_UKIDSS} &
  \multicolumn{1}{c|}{[pmRA, pmDEC]} &
  \multicolumn{1}{c|}{pm}\\
& (deg) & (deg) & (deg) & (deg) & (deg) & (deg) & (mas yr$^{-1}$) & (mas yr$^{-1}$)\\

\noalign{\smallskip}
\hline
\noalign{\smallskip}
  12593776+0651204$^{*}$ & 194.907371 & 6.855669 & 194.907001 & 6.855464 & 194.906650 & 6.855262 & [-372 (24), -218 (12)] & 431 \\
  12593933+0651255 & 194.913892 & 6.857086 & 194.913513 & 6.856873 & 194.913155 & 6.856681 & [-372 (13), -225 (9)] & 434 \\ 
\hline
  12594856+2412318 & 194.952345 & 24.208853 & 194.952236 & 24.208815 & 194.952109 & 24.208791 & [-86 (4), -28 (2)] & 91 \\
  13002787+2412195 & 195.116132 & 24.205439 & 195.116004 & 24.205378 & 195.115911 & 24.205360 & [-83 (1), -30 (2)] & 89 \\
\hline
  13133339+2642534 & 198.389158 & 26.714849 & 198.389012 & 26.714887 & 198.388942 & 26.714966 & [-72 (15), 41 (11)] & 83 \\
  13133381+2642533 & 198.390890 & 26.714827 & 198.390766 & 26.714857 & 198.390700 & 26.714933 & [-62 (11), 38 (9)] & 72 \\
\hline
  13213074+3538472 & 200.378084 & 35.646454 & 200.377977 & 35.646372 & 200.377839 & 35.646380 & [-83 (13), -24 (14)] & 86 \\
  13213079+3538516 & 200.378323 & 35.647667 & 200.378236 & 35.647591 & 200.378078 & 35.647604 & [-83 (13), -21 (14)] & 85 \\
\hline
  13252369+3555344 & 201.348721 & 35.926247 & 201.348529 & 35.926342 & 201.348286 & 35.926386 & [-119 (1), 44 (1)] & 127 \\
  13253949+3604076 & 201.414576 & 36.068779 & 201.414368 & 36.068838 & 201.414160 & 36.068907 & [-130 (3), 38 (0.4)] & 135 \\
\hline
  13261177+3026260 & 201.549077 & 30.440556 & 201.548889 & 30.440601 & 201.548669 & 30.440698 & [-123 (1), 32 (3)] & 127 \\
  13261214+3026235 & 201.550617 & 30.439877 & 201.550429 & 30.439888 & 201.550207 & 30.439987 & [-112 (3), 35 (5)] & 117 \\
\hline
  13371237+1232212 & 204.301547 & 12.539231 & 204.301378 & 12.539181 & 204.301332 & 12.539125 & [-89 (3), -46 (2)] & 100 \\ 
  13373414+1229314 & 204.392287 & 12.492077 & 204.392122 & 12.492042 & 204.392081 & 12.491976 & [-98 (4), -78 (4)] & 126 \\
\hline
  13471881+0746120 & 206.828387 & 7.770011 & 206.828280 & 7.770051 & 206.827927 & 7.770095 & [-154 (2), 32 (2)] & 157 \\
  13471892+0746006 & 206.828842 & 7.766850 & 206.828751 & 7.766906 & 206.828398 & 7.766948 & [-148 (1), 23 (4)] & 149 \\
\hline
  13475983+3343241 & 206.999317 & 33.723373 & 206.999467 & 33.723156 & 206.999667 & 33.722933 & [98 (3), -135 (2)] & 167 \\
  13480028+3341587 & 207.001208 & 33.699646 & 207.001383 & 33.699479 & 207.001618 & 33.699302 & [114 (3), -105 (2)] & 155 \\
\hline
  13490345+1029167 & 207.264378 & 10.487984 & 207.264335 & 10.487902 & 207.264231 & 10.487844 & [-68 (12) , -71 (11)] & 98 \\
  13490709+1030240 & 207.279556 & 10.506668 & 207.279508 & 10.506615 & 207.279419 & 10.506513 & [-72 (2), -80 (2)] & 107 \\
\hline
  13570417+2737490 & 209.267416 & 27.630281 & 209.267544 & 27.630161 & 209.267576 & 27.630020 & [34 (3), -96 (1)] & 102 \\
  13571335+2738194 & 209.305662 & 27.638729 & 209.305761 & 27.638622 & 209.305817 & 27.638445 & [47 (13), -103 (5)] & 113 \\
\hline
  13570535+3403459 & 209.272296 & 34.062771 & 209.272196 & 34.062776 & 209.272032 & 34.062807 & [-82 (13), 13 (6)] & 84 \\
  13570929+3403319 & 209.288749 & 34.058872 & 209.288578 & 34.058909 & 209.288432 & 34.058910 & [-81 (1), 11 (5)] & 82 \\
\hline
  13565606+0552499 & 209.233619 & 5.880534 & 209.233458 & 5.880416 & 209.233464 & 5.880393 & [-47 (4), -63 (3)] & 78 \\
  13572013+0550251 & 209.333897 & 5.840330 & 209.333773 & 5.840233 & 209.333700 & 5.840204 & [-73 (2), -58 (4)] & 94 \\ 
\hline
  16183619+3005336 & 244.650797 & 30.092670 & 244.650696 & 30.092727 & 244.650623 & 30.092792 & [-27 (1), 41 (7)] & 49 \\
  16184068+2958293 & 244.669529 & 29.974827 & 244.669449 & 29.974896 & 244.669312 & 29.974954 & [-31 (4), 44 (3)] & 54 \\
\hline
  16211986+2653207 & 245.332766 & 26.889111 & 245.332698 & 26.889148 & 245.332553 & 26.889213 & [-54 (2), 64 (10)] & 84 \\
  16212591+2649353 & 245.357968 & 26.826477 & 245.357886 & 26.826511 & 245.357677 & 26.826558 & [-73 (10), 66 (15)] & 98 \\
\hline
  16234190+3200293 & 245.924619 & 32.008152 & 245.924567 & 32.008189 & 245.924493 & 32.008230 & [-27 (2), 29 (2)] & 40 \\
  16241587+3155100 & 246.066141 & 31.919455 & 246.066076 & 31.919487 & 246.065998 & 31.919527 & [-36 (3), 33 (2)] & 49 \\
\hline
  22445443+0856474 & 341.226828 & 8.946525 & 341.226618 & 8.946404 & 341.226578 & 8.946368 & [-115 (11), -53 (9)] & 127 \\
  22445848+0901564 & 341.243695 & 9.032349 & 341.243533 & 9.032174 & 341.243495 & 9.032175 & [-100 (9), -49 (2)] & 111 \\
\hline
  22492429+0517137$^{*}$ & 342.351209 & 5.287160 & 342.351456 & 5.287242 & 342.351458 & 5.287250 & [105 (1), 35 (0.05)] & 111 \\
  22492577+0516592 & 342.357384 & 5.283135 & 342.357634 & 5.283218 & 342.357639 & 5.283225 & [90 (1), 41 (0.5)] & 99 \\
\hline
  23253016+1434198 & 351.375679 & 14.572188 & 351.375636 & 14.572115 & 351.375524 & 14.571939 & [-55 (2), -90 (1)] & 106 \\
  23253026+1434102 & 351.376106 & 14.569505 & 351.376034 & 14.569457 & 351.375948 & 14.569287 & [-38 (2), -92 (6)] & 100 \\
\hline
  23280466+0821168 & 352.019434 & 8.354683 & 352.019330 & 8.354569 & 352.019292 & 8.354507 & [-60 (10), -69 (6)] & 92 \\
  23283688+0822011 & 352.153692 & 8.366992 & 352.153608 & 8.366903 & 352.153535 & 8.366825 & [-44 (1), -69 (0.5)] & 82 \\
\hline
  23433880+1304223 & 355.911670 & 13.072880 & 355.911867 & 13.072947 & 355.911894 & 13.072978 & [69 (2), 31 (5)] & 76 \\
  23435793+1256309 & 355.991385 & 12.941942 & 355.991603 & 12.941997 & 355.991618 & 12.942041 & [66 (0.1), 24 (10)] & 71 \\
\noalign{\smallskip}
\hline
\noalign{\smallskip}
\end{tabular}
\end{center}
$^{*}$ previously found in WDS catalogue.\\
$^{+}$ previously found in SLoWPoKES II.\\
\end{flushleft}
\end{table*}

\begin{table*}
\caption[]{Candidate position data for group B. 
2MASS, SDSS and $WISE$.
}
\label{tab:obs2}
\begin{flushleft}
\begin{center}
\scriptsize
\begin{tabular}{lcccccccc}
\noalign{\smallskip}
\hline
\noalign{\smallskip}
  \multicolumn{1}{|c|}{2MASS} &  \multicolumn{1}{c|}{RA\_2MASS} & \multicolumn{1}{c|}{DEC\_2MASS} &
  \multicolumn{1}{c|}{RA\_SDSS} &  \multicolumn{1}{c|}{DEC\_SDSS} &
  \multicolumn{1}{c|}{RA\_$WISE$} &  \multicolumn{1}{c|}{DEC\_$WISE$} &
  \multicolumn{1}{c|}{[pmRA,pmDEC]} &
  \multicolumn{1}{c|}{pm} \\
& (deg) & (deg) & (deg) & (deg) & (deg) & (deg) & (mas yr$^{-1}$) & (mas yr$^{-1}$)\\

\noalign{\smallskip}
\hline
\noalign{\smallskip}
09411179+3315130$^{*}$ & 145.299135 & $+$33.253632 & 145.299562 & $+$33.253221 & 145.300147 & $+$33.252506 & [280 (10), -384 (9)] & 475 \\
09411195+3315060 & 145.299803 & $+$33.251675 & 145.300237 & $+$33.251249 & 145.300833 & $+$33.250586 & [270 (12), -367 (5)] & 456  \\
\hline
09425716+2351200 & 145.738183 & $+$23.855574 & 145.737895 & $+$23.855128 & 145.737601 & $+$23.854804 & [-148 (13), -220 (3)] & 266 \\
09430114+2349173 & 145.754765 & $+$23.821478 & 145.754459 & $+$23.820927 & -$^{1}$ & -$^{1}$ & [-132 (1), -258 (1)] & 290 \\
 & & & &  & & & &  \\
09425716+2351200 & 145.738183 & $+$23.855574 & 145.737895 & $+$23.855128 & 145.737601 & $+$23.854804 & [-148 (13), -220 (3)] & 266 \\
09430132+2349222 & 145.755539 & $+$23.822838 & 145.755267 & $+$23.822411 & 145.754933 & $+$23.822075 & [-135 (12), -204 (23)] & 244 \\
 & & & &  & & & &  \\
09430114+2349173 & 145.754765 & $+$23.821478 & 145.754459 & $+$23.820927 & -$^{1}$ & -$^{1}$ & [-132 (1), -258 (1)] & 290 \\
09430132+2349222 & 145.755539 & $+$23.822838 & 145.755267 & $+$23.822411 & 145.754933 & $+$23.822075 & [-135 (12), -204 (23)] & 244 \\
\hline
10294440+2545374 & 157.435032 & $+$25.760401 & 157.435232 & $+$25.760251 & 157.435376 & $+$25.760146 & [96 (5), -78 (2)] & 124 \\
10294567+2546050 & 157.440331 & $+$25.768072 & 157.440522 & $+$25.767925 & 157.440665 & $+$25.767836 & [100 (7), -78 (4)] & 127 \\
\hline
13034509+2134235$^{*}$ & 195.937905 & $+$21.573198 & 195.937654 & $+$21.573042 & 195.937332 & $+$21.572885 & [-205 (20), -97 (1)] & 227 \\
13034574+2134210 & 195.940608 & $+$21.572527 & 195.940352 & $+$21.572377 & 195.940034 & $+$21.572201 & [-208 (21), -113 (7)] & 237 \\
\hline
13090058+1709066 & 197.252421 & $+$17.151836 & 197.252184 & $+$17.151897 & 197.252010 & $+$17.151946 & [-94 (10), 46 (7)] & 104 \\
13092549+1714584 & 197.356217 & $+$17.249577 & 197.355927 & $+$17.249660 & 197.355776 & $+$17.249696 & [-106 (9), 45 (6)] & 115 \\
\hline
13104398+1434338$^{*}$ & 197.683291 & $+$14.576062 & 197.683245 & $+$14.576131 & 197.683074 & $+$14.576244 & [-81 (8), 66 (2)] & 104 \\
13104431+1434326 & 197.684646 & $+$14.575738 & 197.684587 & $+$14.575803 & 197.684390 & $+$14.575869 & [-82 (8), 50 (2)] & 96 \\
\hline
18163409-1246310$^{*}$ & 274.142048 & $-$12.775305 & 274.141560$^{2}$ & $-$12.775722$^{2}$ & 274.141358 & $-$12.775900 & [-199 (11), -194 (10)] & 278 \\
18163485-1246421 & 274.145216 & $-$12.778377 & 274.144738$^{2}$ & $-$12.778816$^{2}$ & 274.144515 & $-$12.778945 & [-193 (11), -188 (10)] & 269 \\
\noalign{\smallskip}
\hline\noalign{\smallskip}
\end{tabular}
\end{center}
$^{*}$ previously found in WDS catalogue.\\
$^{1}$ Not seen in $WISE$.\\
$^{2}$ GLIPMSE data base coordinates.\\ 
\end{flushleft}
\end{table*}

\begin{table*}
\caption[]{Physical properties of newly identified low-mass component systems: group A. Components ordered by spectral type,
 asignating primary position to the earliest type of the componets.}
\label{tab:para}
\begin{flushleft}
\begin{center}
\scriptsize
\begin{tabular}{cccccccccc}
\noalign{\smallskip}
\hline
\noalign{\smallskip}
Obj. Id. & Spt. type$^{1}$ & $T_{eff}$$^{2}$ & log$g$$^{2}$ &  Masses$^{2}$ & Masses$^{3}$ & Distance & $\rho$ & Separation & U \\
 (2MASS)  & &  (K) & [cm s$^{-2}$] & (M$_{\odot}$) & (M$_{\odot}$) & (pc) & (arcsec) & (AU) & (10$^{33}$ J) \\
\noalign{\smallskip}
\hline
\noalign{\smallskip}
00384499+0519003 & M4.0 & 3300 & 5.5 & 0.20 & 0.24 & 87.53$_{+14}^{-17}$ & 549.23 & 49042 & -0.7\\
00391056+0525351 & M6.5 & 2900 & 5.0 & 0.09 & 0.17 & 91.05$_{+15}^{-18}$ & & & \\ 
\hline
00421418+0431200 & M6.5 & 2900 & 5.0 & 0.10 & 0.17 & 90.79$_{+15}^{-18}$ & 9.03 & 778 & -22.5 \\
00421358+0431185 & M7.5 & 2800 & 5.0 & 0.10 & 0.18 & 81.43$_{+13}^{-16}$ & & & \\
\hline
00513688+0735177 & M5.0 & 3000 & 6.0 & 0.11 & 0.20 & 71.78$_{+12}^{-14}$ & 14.72 & 1110 & -17.6  \\
00513590+0735192 & M5.5 & 3000 & 5.0 & 0.10 & 0.18 & 79.07$_{+13}^{-16}$ & & & \\
\hline
01463893+1545360 & M4.5 & 3100 & 4.5 & 0.11 & 0.20 & 53.41$_{+9}^{-11}$ & 4.79  & 257 & -93.5 \\
01463861+1545371 & M5.0 & 3100 & 5.5 & 0.12 & 0.18 & 54.08$_{+9}^{-11}$ & & & \\
\hline
01560037+0528494 & M7.5 & 2800 & 5.5 & 0.09 & 0.17 & 77.99$_{+13}^{-15}$ & 184.31 & 13338 & -2.4 \\
01554912+0527334 & M8.5 & 2700 & 4.5 &   -  & 0.20 & 66.74$_{+11}^{-13}$ & & & \\ 
 & & & & & & & & &   \\
01560053+0528562 & M6.0 & 2900 & 4.5 & 0.08 & 0.17 & 75.74$_{+12}^{-15}$ & 184.31 & 13338 & -2.2 \\ 
01554912+0527334 & M8.5 & 2700 & 4.5 &   -  & 0.20 & 66.74$_{+11}^{-13}$ & & & \\
 & & & & & & & & &   \\
01560053+0528562 & M6.0 & 2900 & 4.5 & 0.08 & 0.17 & 75.74$_{+12}^{-15}$ & 7.17 & 551 & -24.1 \\
01560037+0528494 & M7.5 & 2800 & 5.5 & 0.09 & 0.17 & 77.99$_{+13}^{-15}$ & & & \\
\hline
01575468+0923422 & M5.5 & 3000 & 4.5 & 0.10 & 0.17 & 96.77$_{+16}^{-19}$ & 10.12 & 986 & -18.4 \\
01575409+0923371 & M6.5 & 2900 & 4.5 & 0.10 & 0.17 & 98.10$_{+16}^{-19}$ & & & \\
\hline
02014517+1124244 & M2.5 & 3300 & 4.5 & 0.21 & 0.38 & 130.66$_{+22}^{-26}$ & 586.25 & 79984 & -0.5 \\
02020000+1115202 & M5.0 & 3000 & 5.0 & 0.10 & 0.18 & 142.20$_{+23}^{-28}$ & & & \\ 
\hline
03165635+0617027 & M7.0 & 2800 & 5.0 & 0.09 & 0.18 & 46.58$_{+8}^{-9}$ & 75.91 & 3525 & -3.9 \\
03165886+0618086 & M7.0 & 2800 & 5.0 & 0.09 & 0.18 & 46.29$_{+8}^{-9}$ & & & \\ 
\hline
08255269+2925169 & M5.0 & 3000 & 5.0 & 0.10 & 0.18 & 98.18$_{+16}^{-19}$ & 16.24 & 1578 & -11.9 \\
08255151+2925220 & M5.5 & 3000 & 5.5 & 0.11 & 0.17 & 96.14$_{+16}^{-19}$ & & & \\
\hline
09045044-0158572 & M3.0 & 3500 & 6.0 & 0.38 & 0.36 & 168.64$_{+28}^{-33}$ & 34.15 & 5645 & -12.9 \\
09045115-0159297 & M6.0 & 3000 & 5.0 & 0.11 & 0.17 & 161.99$_{+27}^{-32}$ & & & \\ 
\hline
09102559+0648533 & M5.0 & 3100 & 5.0 & 0.11 & 0.19 & 183.15$_{+30}^{-36}$ & 369.64 & 65247 & -0.3 \\
09105025+0648111 & M5.5 & 3000 & 5.0 & 0.11 & 0.17 & 169.87$_{+28}^{-34}$ & & & \\ 
\hline
09335192+3237270 & M2.0 & 3700 & 6.0 & 0.51 & 0.50 & 160.53$_{+27}^{-32}$ & 470.30 & 81341 & -1.0 \\
09332493+3232033 & M6.5 (M6$^{a}$) & 2900 & 5.0 & 0.09 & 0.17 & 185.38$_{+31}^{-37}$ & & & \\ 
\hline
09442313+0842231 & M4.5 & 3200 & 4.5 & 0.15 & 0.22 & 154.59$_{+26}^{-31}$ & 7.97 & 1188 & -24.6 \\
09442260+0842235 & M6.0 & 3000 & 5.5 & 0.11 & 0.17 & 143.81$_{+24}^{-29}$ & & & \\
\hline
10132950+1135558 & M3.5 & 3100 & 4.5 & 0.11 & 0.27 & 59.09$_{+10}^{-12}$ & 12.92 & 793 & -23.9 \\
10132882+1136041 & M6.5 & 2900 & 5.5 & 0.10 & 0.17 & 63.67$_{+10}^{-13}$ & & & \\
\hline
10402749-0248092 & M3.5 & 3400 & 6.0 & 0.27 & 0.30 & 260.85$_{+43}^{-52}$ & 3.59 & 975 & -49.5 \\
10402763-0248063 & M5.5 & 3000 & 6.0 & 0.10 & 0.17 & 282.76$_{+47}^{-56}$ & & & \\
\hline
11002544+1445240 & M4.5 & 3100 & 5.5 & 0.11 & 0.20 & 144.13$_{+24}^{-29}$ & 7.58 & 1161 & -19.0 \\
11002578+1445299 & M5.0 & 3100 & 6.0 & 0.11 & 0.19 & 162.04$_{+27}^{-32}$ & & & \\
\hline
11242609+1139504 & M0.5 & 4000 & 6.0 & 0.70 & 0.66 & 181.49$_{+30}^{-36}$ & 234.58 & 4660 & -3.0 \\
11243764+1137085 & M6.5 & 3000 & 6.0 & 0.11 & 0.17 & 215.87$_{+36}^{-43}$ & & & \\
\hline
12050629+1508193 & M3.0 & 3500 & 6.0 & 0.35 & 0.33 & 148.61$_{+25}^{-29}$ & 454.84 & 63814 & 1.0 \\
12043718+1505286 & M6.5 & 2900 & 6.0 & 0.10 & 0.17 & 131.99$_{+22}^{-26}$ & & & \\
\hline
12433001-0105152 & M3.0 & 3400 & 6.0 & 0.29 & 0.35 & 262.19$_{+43}^{-52}$ & 18.33 & 4593  & -14.5 \\
12432885-0105208 & M5.0 & 3100 & 6.0 & 0.13 & 0.18 & 238.91$_{+40}^{-47}$ &  & & \\ 
\noalign{\smallskip}
\hline
\noalign{\smallskip}
\end{tabular}
\end{center}
$^{1}$ From photometry (Sect. 4.2.4).\\
$^{2}$ From VOSA (Sect. 4.2.3). \\
$^{3}$ From Kraus \& Hilenbrand (2007) Table 5 colour-mass relations (Sect. 4.2.3).\\
$^{a}$ From spectroscopy. See Sect. 4.3.2.\\
\end{flushleft}
\end{table*}

\addtocounter{table}{-1}
\begin{table*}
\caption[]{Cont.}
\label{tab:para}
\begin{flushleft}
\begin{center}
\scriptsize
\begin{tabular}{cccccccccc}
\noalign{\smallskip}
\hline
\noalign{\smallskip}
Obj. Id. & Spt. type$^{1}$ & $T_{eff}$$^{2}$ & log$g$$^{2}$ &  Masses$^{2}$ & Masses$^{3}$ & Distance & $\rho$ & Separation & U \\
 (2MASS) & & (K) & [cm s$^{-2}$] & (M$_{\odot}$) & (M$_{\odot}$) & (pc) & (arcsec) & (AU) & (10$^{33}$ J) \\
\noalign{\smallskip}
\hline
\noalign{\smallskip}
12593933+0651255 & M8.0 & 2800 & 4.5 & - & 0.19 & 44.56$_{+7}^{-9}$ & 23.86 & 1110 & -66.5 \\
12593776+0651204 & L0.0 & 2700 & 5.0 & - & 0.21 & 48.45$_{+8}^{-10}$ &  & & \\  
\hline
13002787+2412195 & M1.0 & 3700 & 6.0 & 0.60 & 0.57 & 171.19$_{+28}^{-34}$ & 537.93 & 91852 & -1.5 \\
12594856+2412318 & M5.0 & 3100 & 5.0 & 0.13 & 0.18 & 170.31$_{+28}^{-34}$ & & & \\
\hline
13133381+2642533 & M4.5 & 3100 & 5.0 & 0.11 & 0.20 & 102.31$_{+17}^{-20}$ & 5.57 & 552 & -35.2 \\
13133339+2642534 & M7.5 & 2800 & 5.0 & 0.10 & 0.19 & 95.99$_{+16}^{-19}$ & & & \\
\hline
13213079+3538516 & M3.5 & 3400 & 5.5 & 0.25 & 0.27 & 57.13$_{+9}^{-11}$ & 4.42 & 277 & -287.8 \\
13213074+3538472 & M5.5 & 3600 & 6.0 &   -  & 0.18 & 68.21$_{+11}^{-13}$ & & & \\ 
\hline 
13252369+3555344 & M2.5 & 3500 & 4.5 & 0.40 & 0.43 & 117.86$_{+19}^{-23}$ & 547.79 & 60450 & -0.9 \\
13253949+3604076 & M6.5 & 2800 & 5.0 & 0.07 & 0.17 & 102.84$_{+17}^{-20}$ & & & \\
\hline
13261177+3026260 & M4.5 & 3100 & 5.0 & 0.11 & 0.20 & 226.69$_{+38}^{-45}$ & 5.37 & 1183 & -17.2 \\
13261214+3026235 & M5.5 & 3000 & 5.0 & 0.10 & 0.17 & 213.99$_{+35}^{-42}$ & & & \\ 
\hline
13371237+1232212 & M3.0 & 3400 & 6.0 & 0.28 & 0.32 & 207.56$_{+34}^{-41}$ & 361.27 & 77964 & -0.8 \\
13373414+1229314 & M5.0 & 3100 & 6.0 & 0.13 & 0.18 & 224.05$_{+37}^{-44}$ & & & \\ 
\hline
13471881+0746120 & M5.5 & 3000 & 4.5 & 0.10 & 0.17 & 87.08$_{+14}^{-17}$ & 11.49 & 980 & -18.6 \\
13471892+0746006 & M6.5 & 2900 & 5.5 & 0.10 & 0.17 & 83.37$_{+14}^{-16}$ & & & \\ 
\hline
13475983+3343241 & M2.5 & 3400 & 5.5 & 0.29 & 0.37 & 141.60$_{+23}^{-28}$ & 85.60 & 11523 & -4.6 \\
13480028+3341587 & M5.5 & 3000 & 5.0 & 0.10 & 0.17 & 127.61$_{+21}^{-25}$ & & & \\ 
\hline
13490709+1030240 & M2.5 & 3600 & 6.0 & 0.45 & 0.41 & 212.82$_{+35}^{-42}$ & 86.08 & 19207 & -4.6 \\
13490345+1029167 & M6.0 & 3000 & 5.5 & 0.11 & 0.17 & 233.42$_{+39}^{-46}$ & & & \\ 
\hline
13570417+2737490 & M4.0 & 3200 & 5.0 & 0.15 & 0.25 & 133.59$_{+22}^{-26}$ & 125.71 & 18137 & -1.3 \\
13571335+2738194 & M6.0 & 2900 & 5.0 & 0.09 & 0.17 & 154.95$_{+26}^{-31}$ & & & \\ 
\hline
13570929+3403319 & M3.5 & 3300 & 5.5 & 0.20 & 0.28 & 213.91$_{+35}^{-42}$ & 51.04 & 10084 & -3.5 \\
13570535+3403459 & M7.0 (M6$^{a}$) & 2900 & 5.5 & 0.10 & 0.17 & 181.26$_{+30}^{-36}$ & & & \\ 
\hline
13572013+0550251 & M2.5 & 3500 & 5.5 & 0.39 & 0.41 & 164.68$_{+27}^{-33}$ & 387.18 & 67326 & -1.1 \\
13565606+0552499 & M5.5 & 3000 & 5.5 & 0.11 & 0.17 & 183.09$_{+30}^{-36}$ &  & & \\ 
\hline
16183619+3005336 & M6.5 & 2900 & 5.5 & 0.10 & 0.17 & 163.62$_{+27}^{-32}$ & 428.23 & 75195 &  -0.2 \\
16184068+2958293 & M7.0 & 2800 & 5.5 & 0.09 & 0.18 & 187.56$_{+31}^{-37}$ & & & \\ 
\hline
16211986+2653207 & M5.0 & 3000 & 5.0 & 0.10 & 0.18 & 123.76$_{+20}^{-25}$ & 239.57 & 28622 & -0.6 \\
16212591+2649353 & M7.5 & 2800 & 5.0 & 0.10 & 0.19 & 115.19$_{+19}^{-23}$ & & & \\ 
\hline
16234190+3200293 & M2.0 & 3600 & 4.5 & 0.50 & 0.48 & 166.39$_{+28}^{-33}$ & 537.39 & 83771 & -1.1 \\
16241587+3155100 & M5.5 & 3000 & 6.0 & 0.11 & 0.17 & 145.39$_{+24}^{-29}$ & & & \\ 
\hline
22445848+0901564 & M3.0 & 3500 & 6.0 & 0.35 & 0.33 & 94.35$_{+11}^{-25}$ & 314.73 & 30148 & -2.2 \\
22445443+0856474 & M5.5 & 3000 & 5.5 & 0.11 & 0.17 & 92.09$_{+15}^{-18}$ & & & \\
\hline
22492577+0516592 & M3.5 & 3500 & 6.0 &   -  & 0.28 & 52.86$_{+7}^{-13}$ & 26.46 & 1545 & -33.3 \\  
22492429+0517137 & M5.5 & 3000 & 5.0 & 0.10 & 0.18 & 61.52$_{+10}^{-12}$ & & & \\
\hline
23253026+1434102 & M6.5 & 2900 & 5.5 & 0.09 & 0.17 & 133.00$_{+26}^{-20}$ & 9.77 & 1220 & -27.6 \\
23253016+1434198 & M9.0 & 2700 & 5.0 &   -  & 0.21 & 122.13$_{+20}^{-24}$ & & & \\
\hline
23283688+0822011 & M5.0 & 3000 & 5.0 & 0.10 & 0.18 & 97.24$_{+16}^{-19}$ & 480.24 & 48540 & -0.7 \\  
23280466+0821168 & M8.0 & 2800 & 5.5 &   -  & 0.20 & 104.96$_{+17}^{-21}$ & & & \\
\hline
23433880+1304223 & M3.5 & 3300 & 5.5 & 0.20 & 0.27 & 106.39$_{+18}^{-21}$ & 548.07 & 58713 & -1.2 \\
23435793+1256309 & M8.0 & 2800 & 5.5 &   -  & 0.19 & 107.67$_{+18}^{-22}$ & & & \\ 
\noalign{\smallskip}
\hline
\noalign{\smallskip}
\end{tabular}
\end{center}
$^{1}$ From photometry (Sect. 4.2.4).\\
$^{2}$ From VOSA (Sect. 4.2.3). \\
$^{3}$ From Kraus \& Hilenbrand (2007) Table 5 colour-mass relations (Sect. 4.2.3).\\
$^{a}$ From spectroscopy. See Sect. 4.3.2.\\
\end{flushleft}
\end{table*}

\begin{table*}
\caption[]{Physical properties of newly identified low-mass component systems: group B. Components ordered by spectral type,
 asignating primary position to the earliest type of the componets.}
\label{tab:parab}
\begin{flushleft}
\begin{center}
\scriptsize
\begin{tabular}{cccccccccc}
\noalign{\smallskip}
\hline
\noalign{\smallskip}
Obj. Id. & Spt. type$^{1}$ & $T_{eff}$$^{2}$ & log$g$$^{2}$ &  Masses$^{2}$ & Masses$^{3}$ & Distance & $\rho$ & Separation & U \\
 (2MASS)  & &  (K) & [cm s$^{-2}$] & (M$_{\odot}$) & (M$_{\odot}$) & (pc) & (arcsec) & (AU) & (10$^{33}$ J) \\
\noalign{\smallskip}
\hline
\noalign{\smallskip}
09411195+3315060 & M5.0 & 3500$^{a}$ & 4.5 & -    & 0.18 & 30.10$^{b}$ & 7.44 & 244 & -299.5 \\
09411179+3315130 & L0.0 & 2500 & 4.5 & -    & 0.23 & 35.52 &      &  &  \\
\hline
09430132+2349222 & M0.5 & 3800 & 6.0 & -    & 0.46 & 31.55 & 133.39 & 4640 & -35.0 \\
09425716+2351200 & M8.5 (M7$^{c}$) & 2700 & 4.5 & -    & 0.20 & 34.12 &        &  &  \\
 & & & & & & & & &   \\
09430114+2349173 & M7.0 & 2700 & 5.5 & 0.06 & 0.14$^{d}$ & 38.68$^{e}$ & 134.10 & 4715 & -10.0 \\
09425716+2351200 & M8.5 (M7$^{c}$) & 2700 & 4.5 & -    & 0.20 & 34.12 &        &  &  \\
 & & & & & & & & &   \\
09430132+2349222 & M0.5 & 3800 & 6.0 & -    & 0.46 & 31.55 & 5.6 & 195 & -249.8  \\
09430114+2349173 & M7.0 & 2700 & 5.5 & 0.06 & 0.14$^{d}$ & 38.68$^{e}$ & &  &  \\
\hline
10294440+2545374 & M4.0 & 3200 & 4.5 & 0.15 & 0.18 & 26.77 & 33.56 & 950 & -19.5 \\
10294567+2546050 & M6.5 & 2800 & 4.5 & 0.07 & 0.17 & 29.82 &       &  &  \\
\hline
13034509+2134235 & M6.0 & 2900 & 4.5 & 0.11 & 0.17 & 61.59 & 10.03 & 605 & -28.9 \\
13034574+2134210 & M8.0 & 2700 & 5.5 & 0.09 & 0.19 & 58.99 &       &  &  \\
\hline
13092549+1714584 & M6.5 (M6$^{c}$) & 2800 & 5.0 & 0.07 & 0.17 & 74.39 & 513.26 & 39878 & -0.22 \\
13090058+1709066 & M7.5 (M7$^{c}$) & 2800 & 5.5 & 0.07 & 0.18 & 81.00 &        &  &  \\
\hline
13104431+1434326 & M7.0 & 2800 & 4.5 & -    & 0.17 & 65.68 & 5.01 & 333 & -162.2 \\
13104398+1434338 & M7.5 & 2800 & 4.5 & -    & 0.18 & 67.32 &      &  &  \\
\hline
18163409-1246310 & M2.0$^{f}$ (M4-M6$^{c}$) & 3500 & 5.0 & -    & 0.45$^{g}$ & 42.27$^{h}$ & 15.85 & 698 & -182.1 \\
18163485-1246421 & M4.5$^{f}$ (M5-M6$^{c}$) & 3100 & 5.5 & -    & 0.16$^{g}$ & 45.79$^{h}$ &       &  &  \\
\noalign{\smallskip}
\hline
\noalign{\smallskip}
\end{tabular}
\end{center}
$^{1}$ From photometry (Sect. 4.2.4).\\
$^{2}$ From VOSA (Sect. 4.2.3). \\
$^{3}$ From Kraus \& Hilenbrand (2007) Table 5 colour-mass relations (Sect. 4.2.3).\\
$^{a}$ No good fit of the SED was found, 3500-3600 K was the best but we take his value with caution,
 relying more in the temperature obtained by other methods. \\
$^{b}$ From $M_r-(i-z)$ relation in Bochanski et al. (2010; Table 4), since the (r-z) colour is outside the allowed range for the $Mr - (r-z)$ calibration. See Sect. 4.2.1.\\
$^{c}$ From spectroscopy. See Sect. 4.3.2.\\
$^{d}$ From $M_i$ and $M_r$ since $z$ value is not reliable.\\
$^{e}$ From $M_r-(r-i)$ relation in Bochanski et al. (2010; Table 4), since z value is not reliable. See Sect. 4.2.1.\\
$^{f}$ From $(H-K_s)$ colour (Sect. 4.2.4).\\
$^{g}$ From $M_J$ relations in Kraus \& Hilenbrand (2007). See Sect. 4.2.3.\\
$^{h}$ From $M_J$-spectral type relation in Hawley et al. (2002) (Sect. 4.2.4).\\
\end{flushleft}
\end{table*}


\begin{table*}
\caption[]{Physical properties of the identified higher mass tertiaries to the M-M systems}
\label{tab:paratertiaries}
\begin{flushleft}
\begin{center}
\scriptsize
\begin{tabular}{cccccccccccc}
\noalign{\smallskip}
\hline
\noalign{\smallskip}
System Id. & PPMXL ID. & {RA} & {DEC} & {[pmRA, pmDEC]} & Spt. Type & $T_{eff}$ & Masses & Distance & Separation & U \\
 (2MASS) & & (deg) & (deg) & (mas yr$^{-1}$) & & (K) & M$_{\odot}$ & (pc) & (AU) & (10$^{33}$ J) \\
\noalign{\smallskip}
\hline
\noalign{\smallskip}
02014517+1124244 & 2102675187759530038 &  29.854309  &  11.281106 &  [65 (1.6),-32 (1.6)]  & G7 & 5500 & 0.90 & 148.08 & 275725.81 &  1.8 \\ 
02020000+1115202 &                     &             &            &            &    &  &      &       &           &     \\ 
\hline
09335192+3237270 & 4233992647964279227 &  143.245865 &  32.478501 &  [-61 (4.8),-38 (4.8)] & K6 & 4400 & 0.64 & 163.30 & 136476.81 &  5.0 \\ 
09332493+3232033 &                     &             &            &            &    &  &      &       &           &     \\
                       &                     &             &            &          &  &      &      &       &           &     \\
09335192+3237270 & 4234014784472339948 &  143.566876 &  32.607488 &  [-51 (2.0),-49 (2.1)] & K0 & 5200 & 0.80 & 189.97 & 49867.50  &  17.0 \\ 
09332493+3232033 &    &             &            &            &    &  &      &       &           &     \\
\hline
11242609+1139504 & 4137926243272078734 &  171.108659 &  11.664022 &  [75 (5.3),-72 (5.3)]  & K7 & 4300 & 0.61 & 156.89 & 50676.59  &  17.2 \\ 
11243764+1137085 &                     &             &            &            &    &  &      &       &           &     \\
\hline
13475983+3343241 & 4551655942546822025 &  207.003151 &  33.742283 &  [78 (3.8),-135 (3.8)] & K9 & 3900 & 0.60 & 141.65 & 9775.82   &  42.2 \\ 
13480028+3341587 &                     &             &            &            &    &  &      &       &           &     \\
\hline 
13572013+0550251 & 4398878071445995714 &  209.242262 &  5.856395  &  [-37 (4.6),-40 (4.6)] & K0 & 5200 & 0.80 & 164.47 & 54876.92  &  12.9 \\ 
13565606+0552499 &      &             &            &            &   &   &      &       &           &     \\
\noalign{\smallskip}
\hline
\noalign{\smallskip}
\end{tabular}
\end{center}
\end{flushleft}
\end{table*}

\begin{table*}
\caption[]{Physical properties of the new L-F system}
\label{tab:paraLbinaries}
\begin{flushleft}
\begin{center}
\scriptsize
\begin{tabular}{ccccccccc}
\noalign{\smallskip}
\hline
\noalign{\smallskip}
Obj. Id. & [pmRA,pmDEC]/pm & Spt. type & $T_{eff}$$^{1}$ & log$g$$^{1}$ & Masses & Distance & Separation & U \\
         & (mas yr$^{-1}$)  &  & (K) & [cm s$^{-2}$] & (M$_{\odot}$) & (pc) & (AU) & (10$^{33}$ J) \\
\noalign{\smallskip}
\hline
\noalign{\smallskip}
2MASS 09320299+1231027    & [-125 (7.1), -113 (7.1)]/169 & L2.0$^{a}$ & 2300.0 & 5.5 & 0.075$^{b}$ & 40 & 130450 & 1.2\\
PPMXL 4077732287929300487 & [-111 (1.0), -125 (0.6)]/167 & F8.0$^{c}$ & 6100.0 &     & 1.19$^{c}$  & 52 &        &    \\
\noalign{\smallskip}
\hline
\noalign{\smallskip}
\end{tabular}
\end{center}
$^{1}$ From VOSA (Sect. 4.2.3).\\
$^{a}$ From photometry (Sect. 4.2.4).\\
$^{b}$ Value state by default for an L dwarf. See Sect 6.\\
$^{c}$ From temperature and spectral types in Gray (1992). See Sect 6.\\ 
\end{flushleft}
\end{table*}


\section*{Acknowledgments}
 M.C. G\'alvez-Ortiz acknowledges the financial support of a JAE-Doc CSIC fellowship 
 co-funded with the European Social Fund under the 
 programme {\em ”Junta para la Ampliaci\'on de Estudios”} and
 the support of the Spanish Ministry of Economy and Competitiveness (MINECO) through the project
 AYA2011-30147-C03-03 and AYA2014-54348-C3-2-R.
 This publication makes use of VOSA, developed under the Spanish Virtual 
 Observatory project supported from the Spanish ministry of Education and Competitiveness
 (MINECO) through grant AYA2014-55216.
 N. L. was funded by the Ram\'on y Cajal fellowship number 08-303-01-02 
 and his research partially funded by the Spanish ministry of Education
 and Competitiveness (MINECO) under programmes
 AYA2010-19136 and AYA2015-69350-C3-2-P.

 This work used observations made with LIRIS on William Herschel Telescope 
 operated on the island of La Palma by The Isaac Newton Group of
 Telescopes (ING), and with OSIRIS at the Gran Telescopio Canarias (GTC)
 programme number GTC38\_15A,
 both in the Spanish Observatorio del Roque de los Muchachos 
 of the Instituto de Astrof\'isica de Canarias.

 Also based on observations collected at the Centro Astron\'omico Hispano Alem\'an (CAHA) 
 at Calar Alto, operated jointly by the Max-Planck Institut f$\ddot{u}$r Astronomie and the 
 Instituto de Astrof\'isica de Andaluc\'ia (CSIC)


\bsp
\label{lastpage}

\appendix
\section{Tables}

\begin{table*}
\caption[]{L-T Candidates.
}
\label{tab:obsL}
\begin{flushleft}
\begin{center}
\tiny
\begin{tabular}{lccccccccccc}
\noalign{\smallskip}
\hline
\noalign{\smallskip}
  \multicolumn{1}{|c|}{2MASS} &  \multicolumn{1}{c|}{RA\_2MASS} & \multicolumn{1}{c|}{DEC\_2MASS} &
  \multicolumn{1}{c|}{RA\_SDSS} &  \multicolumn{1}{c|}{DEC\_SDSS} & $2MASS\_J$ &
  \multicolumn{1}{c|}{[pmRA,pmDEC]} &
  \multicolumn{1}{c|}{pm} & Spt. type$^{1}$ & $T_{eff}$$^{2}$ & log$g$$^{2}$ & Dist.\\
& (deg) & (deg) & (deg) & (deg) & (2MASS) & (mas yr$^{-1}$) & (mas yr$^{-1}$) & & (K) & [cm s$^{-2}$] & (pc) \\
\noalign{\smallskip}
\hline
\noalign{\smallskip}
00062250+1300451  &  1.593781  &  13.012551  &  1.593775  &  13.012333  & 16.96 &  [-2, -69]  &  69  & L2.0 & 1800 &  5.5  &  78 \\
00081285+0806441  &  2.053555  &  8.11226  &  2.053566  &  8.111889  & 16.59 & [8, -270]  &  270  & L1.5 & 2000 &  4.5  &  70 \\
00363231+0722108  &  9.134658  &  7.369681  &  9.134572  &  7.369572  & 16.55 &  [-61, -78]  &  99  & L1.5 & 2500 &  4.5  &  72 \\
01250319+0840499  &  21.263297  &  8.680554  &  21.263427  &  8.680479  & 16.47 & [57, -33]  &  66  & L2.0 & 1600 &  4.5  &  57 \\
02151451+0453179  &  33.810475  &  4.888312  &  33.810537  &  4.887977  & 16.60 & [27, -148]  &  151  & L2.5 & 1500 &  5.0  &  65 \\
03074939+0516533  &  46.955794  &  5.281497  &  46.955758  &  5.28135  & 16.76 & [-30, -123]  &  127  & L2.5 & 1500 &  5.0  &  57 \\
07415323+2129056  &  115.471829  &  21.484909  &  115.471803  &  21.485079  & 16.64 & [-18, 124]  &  125  & L2.5 & 2300 &  4.5  &  75 \\
08064841+2215456  &  121.701749  &  22.262667  &  121.701728  &  22.262519  & 16.99 & [-22, -171]  &  172  & L1.0$^{a}$ & 1700 &  4.5  &  94$^{a}$ \\
08223562+0442042  &  125.64844  &  4.701173  &  125.648228  &  4.701246  & 16.75 & [-372, 128]  &  393  & L5.0 & 1800 &  5.0  &  33 \\
08330964+2949094  &  128.290178  &  29.819294  &  128.29016  &  29.819006  & 16.34 & [-11, -211]  &  211  & L2.5 & 1600 &  5.0  &  49 \\
08532128-0117039  &  133.33867  &  -1.284424  &  133.338385  &  -1.284338  & 16.69 & [-125, 38]  &  130  & L2.0 & 2400 &  4.5  &  68 \\
09165461+0546086  &  139.227582  &  5.769081  &  139.227618  &  5.768987  & 16.24 & [63, -166]  &  178  & L2.0 & 1500 &  5.5  &  53 \\
09320299+1231027  &  143.012464  &  12.517433  &  143.012209  &  12.517207  & 15.60 & [-125, -113]  &  169  & L2.0 & 2300 &  5.5  &  40 \\
09412866+0504166  &  145.369443  &  5.071281  &  145.36924  &  5.071271  & 16.37 & [-102, -5]  &  102  & L1.5 & 2500 &  4.5  &  59 \\
09443669+3356113  &  146.152891  &  33.936493  &  146.152801  &  33.936618  & 16.87 & [-45, 76]  &  88  & L2.5 & 2400 &  4.5  &  61 \\
09461127+3027152  &  146.546992  &  30.454231  &  146.546882  &  30.454104  & 16.74 & [-57, -76]  &  95  & L2.0 & 2400 &  4.5  &  76 \\
10355745+1149420  &  158.98938  &  11.828347  &  158.989305  &  11.828318  & 16.12 & [-95, -37]  &  102  & L1.5$^{b}$  & 2200 &  5.5  &  57 \\
10505470-0048352  &  162.727918  &  -0.809791  &  162.727382  &  -0.809848  & 16.17 & [-234, -25]  &  235  & L1.5$^{b}$ & 2500 &  5.0  &  59 \\
10520456+0617307  &  163.019035  &  6.291863  &  163.018792  &  6.291908  & 16.87 & [-445, 83]  &  452  & L3.5 & 2000 &  5.5  &  55 \\
10580444+1339474** &  164.518513  &  13.663172  &  164.517942  &  13.663039  & 16.43 & [-332, -79]  &  341  & L2.5 & 2400 &  5.5  &  49 \\
11025112+1040469  &  165.713021  &  10.679718  &  165.712878  &  10.679678  & 16.34 & [-166, -47]  &  172  & L2.5 & 2500 &  4.5  &  57 \\
11072127+1754418  &  166.838629  &  17.911617  &  166.838517  &  17.911491  & 16.76 & [-52, -61]  &  80  & L1.5 & 1800 &  5.5  &  72 \\
11082937+1543012  &  167.122404  &  15.717017  &  167.122135  &  15.717025  & 16.69 & [-125, 4]  &  125  & T0-T2$^{c}$ & 1800 &  5.0  &  -$^{c}$ \\
11220855+0343193  &  170.535636  &  3.722044  &  170.535561  &  3.722026  & 15.65 & [-231, -56]  &  238  & L3.0 & 2300 &  4.5  &  38 \\
11243866+1542585  &  171.161099  &  15.716263  &  171.160612  &  15.716392  & 16.70 & [-283, 78]  &  293  & L2.0 & 2500 &  5.5  &  68 \\
11305248+1638019** &  172.718694  &  16.633886  &  172.718339  &  16.633741  & 16.46 & [-199, -85]  &  216  & L1.5 & 2500 &  5.0  &  70 \\
11411784+0108184  &  175.324349  &  1.138468  &  175.324255  &  1.138331  & 16.27 & ---  &  ---  &  L2.0  & 2400 &  5.5  &  56 \\
12051260+1245381  &  181.30254  &  12.760606  &  181.302397  &  12.760703  & 16.39 & [-96, 67]  &  117  & L2.0 & 2400 &  4.5  &  56 \\
12594167+1001380  &  194.923656  &  10.027239  &  194.923584  &  10.027265  & 16.82 & [-130, 48]  &  138 & L4.0 & 1700 &  5.0  &  54 \\
13023109+2648294  &  195.629559  &  26.808191  &  195.629383  &  26.808316  & 16.79 & [-99, 79]  &  127  & L2.5 & 1800 &  5.5  &  62 \\
13084263+0432441* &  197.177664  &  4.54559  &  197.177768  &  4.545449  & 16.07 & [341, -463]  &  575  & L1.5 & 2500 &  5.0  &  56 \\
13211687+2755329  &  200.320319  &  27.925825  &  200.320568  &  27.925768  & 16.81 & [157,-41]  &  162  & L1.5 & 2200 &  5.5  &  89 \\
13284086+0746280  &  202.170254  &  7.774453  &  202.169892  &  7.774319  & 16.10 & [-209, -78]  &  223  & L2.0 & 2400 &  4.5  &  51 \\
13364870+2134050  &  204.20292  &  21.568056  &  204.202661  &  21.568029  & 16.88 & [-147, -16]  &  148  & L2.5 & 2200 &  5.5  &  67 \\
13493145+2945533  &  207.38108  &  29.764826  &  207.381004  &  29.764848  & 15.77 & [-58,19]  &  061  &  L2.0  &  1800.0  &  5.5  &  43 \\
13510003+2925273  &  207.750126  &  29.424257  &  207.749752  &  29.424225  & 16.97 & [-291, -28]  &  292  & T0-T2$^{c}$ & 1200 &  4.5  &  -$^{c}$ \\
14025073+3639378  &  210.711385  &  36.660507  &  210.711031  &  36.660493  & 16.98 & [-204, -10]  &  204  & L2.5 & 2000 &  5.5  &  43 \\
14092137+0818363  &  212.339044  &  8.310092  &  212.339011  &  8.309978  & 16.09 & [-38, -133]  &  139  & L1.5 & 1800 &  5.0  &  90 \\
14120397+1216100  &  213.016546  &  12.269448  &  213.016283  &  12.269446  & 16.39 & [-268, -2]  &  268  & L4.0 & 1900 &  4.5  &  34 \\
14193426+1413257  &  214.892778  &  14.223807  &  214.892742  &  14.223641  & 16.54 & [-20, -95]  &  97  & L1.5 & 2400 &  4.5  &  70 \\
14313545-0313117  &  217.897746  &  -3.219942  &  217.897317  &  -3.219785  & 16.09 & [-189, 69]  &  202  & L5.0 & 1800 &  5.0  &  23 \\
15210289-0008348  &  230.262079  &  -0.143  &  230.262281  &  -0.143074  & 16.66 &  ---  &  ---  & L2.0 & 1800 &  4.5  &  70 \\
15543602+2724487  &  238.650105  &  27.413544  &  238.650176  &  27.413418  & 16.19 & [72, -145]  &  162  & L4.0 & 1800 &  5.5  &  31 \\
20595810-0012356  &  314.992101  &  -0.209907  &  314.992026  &  -0.210154  & 16.74 & [-53, -175]  &  183  & L2.5 & 1800 &  5.0  &  68 \\
22195949+0451337  &  334.997876  &  4.859363  &  334.998038  &  4.859271  & 16.35 & [70, -40]  &  81  & L2.0 & 2400 &  5.5  &  59 \\
22355244+0418563  &  338.968505  &  4.31564  &  338.967821  &  4.315269  & 15.37 & [-301, -163]  &  342  & L4.5 & 1800 &  5.5  &  20 \\
22525796+0209332  &  343.241504  &  2.159227  &  343.241795  &  2.158899  & 16.61 & [129, -145]  &  194  & L1.5 & 2200 &  4.5  &  69 \\
23004298+0200145  &  345.179123  &  2.004042  &  345.179428  &  2.004025  & 16.40 & [136, -8]  &  136  & L3.5 & 1700 &  4.5  &  53 \\
23132142+0253472  &  348.339288  &  2.896458  &  348.338946  &  2.896158  & 16.80 & [-152, -133]  &  202  & L2.0 & 2300 &  5.0  &  67 \\
23292520+1009302  &  352.355018  &  10.158415  &  352.354921  &  10.158206  & 16.45 & [-42, -93]  &  102  & L2.0 & 2300 &  4.5  &  56 \\
\noalign{\smallskip}
\hline
\noalign{\smallskip}
\end{tabular}
\end{center}
$^{1}$ From photometry (Sect. 4.2.4).\\
$^{2}$ From VOSA (Sect. 4.2.3). \\
$^{a}$ From ($i-J$) colour-spectral type relation of Schmidt et al. (2010) only since ($i-z$) criterion is not fullfilled.\\
$^{b}$ Classified as L dwarf by SDSS pipeline.\\
$^{c}$ Since ($i-z$) and ($i-J$) criteria of Schmidt et al. (2010) are not fullfilled, the spectral type is calculated
  from the rest of the colours in Table~3 of Schmidt et al. (2010).\\ 
$^{*}$ These objects may correspond to low metallicity Ls, or L subdwarfs, due to their
 proper motion and colours characteristics, see Sect. 5.\\
$^{**}$ Spectrum of the target is available and discussed in Sect. 5.\\
\end{flushleft}
\end{table*}

\begin{table*}
\caption[]{Wide low mass Binaries from the literature. Separations over 200 AU and total system masses under 1 M$\odot$.}
\label{tab:wlmbfl}
\begin{flushleft}
\begin{center}
\tiny
\begin{tabular}{lccccl}
\noalign{\smallskip}
\hline
\noalign{\smallskip}
Name$^{1}$ & Separation & Masses$_{P}$ & Masses$_{S}$$^{2}$ &  U & References\\
   & (AU) &  (M$_{\odot}$) & (M$_{\odot}$) & (10$^{33}$ J) & \\
\noalign{\smallskip}
\hline
\noalign{\smallskip}
2M1043+1706 AB  & 1020 & 0.21  & 0.08 & -29 & Baron et al. (2015)\\
LSPM J1021+3704 \&  2M1021+3704  & 3000 & 0.21  & 0.07 & -8.6  & Baron et al. (2015)\\
LSPM J1236+3000 \&  2M1236+3000  & 1580 & 0.10  & 0.08 & -8.9  & Baron et al. (2015)\\
LSPM J1259+1001 \&  2M1259+1001  & 345  & 0.12  & 0.06 & -36.7 & Baron et al. (2015)\\
LSPM J1441+1856 \&  2M1441+1856  & 4110 & 0.10  & 0.07 & -3.0  & Baron et al. (2015)\\
NLTT 182        \&  2M0005+0626  & 400  & 0.16  & 0.08 & -56.3 & Baron et al. (2015)\\
NLTT 251        \&  2M0006-0852  & 850  & 0.10  & 0.08 & -16.6 & Baron et al. (2015)\\
NLTT 687   \&       2M0013-1816  & 7400 & 0.39  & 0.07 & -6.5  & Baron et al. (2015)\\
NLTT 2274       \&  2M0041+1341  & 725  & 0.21  & 0.08 & -40.8 & Baron et al. (2015)\\
NLTT 20640      \&  2M0858+2710  & 780  & 0.21  & 0.07 & -33.1 & Baron et al. (2015)\\
NLTT 26746 \&       2M1115+1607  & 660  & 0.21  & 0.06 & -33.6 & Baron et al. (2015)\\
NLTT 29392 \&       2M1202+4204  & 310  & 0.10  & 0.07 & -39.7 & Baron et al. (2015)\\
NLTT 30510 \&       2M1222+3643  & 1635 & 0.38  & 0.07 & -28.6 & Baron et al. (2015)\\
NLTT 36369 \&       2M1408+3708  & 590  & 0.21  & 0.09 & -56.4 & Baron et al. (2015)\\
HIP 49046 AB   & 4720 &  0.60 & 0.09  & -20.1 & Deacon et al. (2014)\\
HIP 60501 AB   & 2100 &  0.60 & 0.14  & -70.4 & Deacon et al. (2014)\\
HIP 63506 AB   & 5640 &  0.60 & 0.06  & -11.2 & Deacon et al. (2014)\\
HIP 73169 AB   & 796  &  0.60 & L2.5  & -     & Deacon et al. (2014)\\
HIP 78184 AB   & 3829 &  0.60 & 0.075 & -20.6 & Deacon et al. (2014)\\
LSPM J1336+2541 AB & 8793 & 0.36 & L4   & -     & Deacon et al. (2014)\\
LSPM J2153+1157 AB & 408  & 0.36 & 0.09 & -139.8 & Deacon et al. (2014)\\
NLTT 730 AB        & 5070 & 0.20  & L7.5 & -     & Deacon et al. (2014)\\
NLTT 8245 AB       & 562  & 0.60  & 0.09 & -169.1 & Deacon et al. (2014)\\
NLTT 18587 AB      & 12200 & 0.44 & 0.08 & -5.1  & Deacon et al. (2014)\\
NLTT 19109 AB      & 362  & 0.20  & 0.08 & -77.8  & Deacon et al. (2014)\\
NLTT 22073 AB      & 776  & 0.44 & 0.08 & -79.8  & Deacon et al. (2014)\\
NLTT 26746 AB   & 661  & 0.20  & L4  & - & Deacon et al. (2014)\\
NLTT 27966 AB   & 630  & 0.14 & L4  & - & Deacon et al. (2014)\\
NLTT 29395 AB   & 671  & 0.36 & 0.08 & -75.5  & Deacon et al. (2014)\\
NLTT 30510 AB   & 962  & 0.44 & 0.07 & -56.3  & Deacon et al. (2014)\\
NLTT 31450 AB   & 487  & 0.20  & L6  & - & Deacon et al. (2014)\\
NLTT 38489 AB   & 418  & 0.20  & 0.075 & -63.1  & Deacon et al. (2014)\\
NLTT 39312 AB   & 713  & 0.44 & 0.08 & -86.9  & Deacon et al. (2014)\\
NLTT 44368 AB   & 7760 & 0.36 & L1.5 & - & Deacon et al. (2014)\\
NLTT 52268 AB   & 549  & 0.36 & 0.075 & -86.6  & Deacon et al. (2014)\\
NLTT 55219 AB   & 432  & 0.44 & L5.5 & - & Deacon et al. (2014)\\
PMI 13410+0542 AB & 484 & 0.49 & L4 & - & Deacon et al. (2014)\\
PMI 13518+4157 AB & 613 & 0.40  & L1.5 & - & Deacon et al. (2014)\\
PMI 22118−1005 AB  & 8892 & 0.44 & L1.5 & - & Deacon et al. (2014)\\
PMI 23492+3458 AB  & 949  & 0.44 & L9 & - & Deacon et al. (2014)\\
LHS 6176 \& ULAS J095047.28+011734.3     & 1400 & 0.28 & T8 & - & Luhman et al. (2012)\\
G259-20  \& 2MASS J17430860+8526594      & 650  & 0.40 & L5 & - & H$\o$g et al. (2000), Luhman et al. (2012)\\
LHS 3421   \& 2MASS J18525777−570814     & 1500 & 0.40 & - & - & van Leeuwen (2007), Hawley et al. (1996), Luhman et al. (2012)\\
LSPM J2010+0632 \& 2MASS J20103539+0634367 & 2100 & 0.28 & 0.08 & -18.7 & Luhman et al. (2012)\\
2MASS J05254550-7425263 \& 2MASS J05253876-7426008 & 2020 & 0.46 & 0.07 & -28.1 & Mu$\breve{z}$i\'c et al. (2012)\\
2MASS J13480721-1344321 \& 2MASS J13480290-1344071 & 1400 & 0.18 & 0.04 & -9.1 & Mu$\breve{z}$i\'c et al. (2012)\\
NLTT 2274 \& SDSS J004154.54+134135.5 & 483  & 0.20  & 0.06 & -43.7 & Faherty et al. (2010)\\
G73-26 \& SDSS J020735.60+135556.3   & 2774 & 0.44 & L2 & - & Faherty et al. (2010)\\
G121-42 \& 2MASS J12003292+2048513   & 5916 & 0.20  & 0.09 & -5.3 & Faherty et al. (2010)\\
G204-39 \& SDSS J175805.46+463311.9  & 2685 & 0.36 & 0.02 & -4.8 & Faherty et al. (2010)\\
LP 213-67$^{a}$  & 230 & 0.10 & 0.18 & -138.1  & Gizis et al (2000), Close et al. (2003), Faherty et al. (2010)\\
Wolf 940 \& ULAS 2146 & 400 & 0.20 & 0.03 & -26.5 & Burningham et al. (2009), Faherty et al. (2010)\\
LP 261-75 \& 2MASS J09510549+3558021 & 450 & 0.22 & 0.02 & -17.2 &  Reid \& Walkowicz (2006), Faherty et al. (2010)\\
Gl 618.1 AB  & 1090 & 0.67 & 0.06 & -65.1 & Wilson et al. (2001), Faherty et al. (2010)\\
G124-62 \& DENIS-P J1441-0945 & 1496 & 0.21 & 0.07 & -17.3 & Seifahrt et al. (2005), Faherty et al. (2010)\\
2MASS J12583501+4013083 \& 2MASS J12583798+4014017 & 6700 & 0.11 & 0.09 & -26.1 & Radigan et al. (2009), Faherty et al. (2010)\\
FU Tau AB      & 800  & 0.05  & 0.015 & -1.6 & Luhman et al. (2009)\\
2M0126–50 AB  & 5100 & 0.095 & 0.092 & –3.0 & Caballero (2009), Artigau et al. (2007)\\
2M1258+40 AB  & 6700 & 0.105 & 0.091 & –2.5 & Caballero (2009), Radigan et al. (2009)\\
AU Mic \& AT Mic AB & 46400    &  0.45 & 0.52 & -7.7  & Caballero (2009)\\
LEHPM 494 \& DE 0021􏰈42 (Kö1 AB) & 1800 & 0.103 & 0.079 & –8.0 & Caballero (2007a,b) \& (2009)\\
LP 655-23 \& 2M 0430-08 (Kö2 AB) & 450  & 0.26 & 0.086 & -87.7 & Caballero (2007b)\\
SE70 \& SOri68 & 1700 & 0.045 & 0.005 & -0.2 & Caballero et al. (2006)\\
DENIS J055146.0-443412.2 AB & 220 & 0.085 & 0.079 & -53.5 & Burgasser et al. (2007)\\
2MASS J11011926-7732383 AB  & 242 & 0.05  & 0.025 &  -9.1 & Siegler et al. (2005), Burgasser et al. (2007)\\
Gliese  150.1 AB & 2400 & 0.57 & 0.46 &  -191.7 & Fischer \& Marcy (1992)\\
Gliese  277 AB  &  550  & 0.39 & 0.27 &  -335.9 & Fischer \& Marcy (1992)\\
Gliese  589 AB  &  240  & 0.25 & 0.11 &  -201.0 & Fischer \& Marcy (1992)\\
Gliese  644 AC  &  1730 & 0.33 & 0.09 &  -30.1  & Fischer \& Marcy (1992)\\
Gliese  669 AB  &  210  & 0.30 & 0.20 &  -501.3 & Fischer \& Marcy (1992)\\
Gliese  720 AB  &  2075 & 0.53 & 0.22 &  -98.6  & Fischer \& Marcy (1992)\\
Gliese  745 AB  &  1300 & 0.27 & 0.27 &  -98.4  & Fischer \& Marcy (1992)\\
Gliese  752 AB  &  545  & 0.42 & 0.08 &  -108.2 & Fischer \& Marcy (1992)\\
\noalign{\smallskip}
\hline
\noalign{\smallskip}
\end{tabular}
\end{center}
$^{1}$ As appeared in reference paper.\\
$^{2}$ For spectral types later than L0 without mass determination in the literature, no mass has been asigned.\\  
$^{a}$ B component is a binary M8+L0. Its total mass is larger than primary M6.4 mass.\\
\end{flushleft}
\end{table*}

\begin{table*}
\caption[]{Wide low mass Binaries from the literature (cont.): SLoWPoKES I. }
\label{tab:slow}
\begin{flushleft}
\begin{center}
\tiny
\begin{tabular}{lccc}
\noalign{\smallskip}
\hline
\noalign{\smallskip}
Name   & Separation & Total Mass$^{1}$ &  U$^{1}$ \\
SLoWPoKES  & (AU) & (M$_{\odot}$) & (10$^{33}$ J) \\
\noalign{\smallskip}
\hline
\noalign{\smallskip}
SLW0020-15 & 1324  & 0.69 & -329.3\\
SLW0250+19 & 26411 & 0.35 & -2.7\\
SLW1521+26 & 15772 & 0.39 & -6.5\\
SLW1540+13 & 13983 & 0.35 & -4.5\\
SLW0721+35 & 2714  & 0.35 & -22.3\\
SLW0903+06 & 7330  & 0.35 &  -7.9\\
SLW0935+51 & 3564  & 0.75 & -124.4\\
SLW1204+19 & 994   & 0.30 & -32.9\\
SLW0236-01 & 16955 & 0.34 & -43.9\\
SLW0903+53 & 3326  & 0.43 & -44.9\\
SLW1318+47 & 3031  & 0.39 & -28.8 \\
SLW1410+01 & 1299  & 0.50 & -171.9 \\
SLW1542+50 & 6125  & 0.29 & -4.3\\
SLW1027+49 & 1960  & 0.31 & -22.9\\
SLW1116+05 & 2618  & 0.42 & -45.5\\
SLW1146+43 & 1589  & 0.43 & -99.9\\
SLW1259+47 & 902   & 0.39 & -94.5\\
SLW1121+58 & 1362  & 0.32 & -50.4\\
SLW0336-05 & 440   & 0.30 & -93.3\\
SLW1417+13 & 1211  & 0.39 & -68.7\\
SLW1242+24 & 1561  & 0.30 & -23.6\\
SLW0820+56 & 510   & 0.39 & -268.2\\
SLW0105+15 & 538   & 0.31 & -107.6\\
SLW1259+19 & 1975  & 0.24 & -10.5\\
SLW1840+42 & 870   & 0.21 & -22.8\\
\noalign{\smallskip}
\hline
\noalign{\smallskip}
\end{tabular}
\end{center}
$^{1}$ Binding energies are calculated using estimated masses as a function 
 of spectral type (Kraus \& Hillenbrand 2007). When spectral type not available, 
 it was assumed to be an equal-mass binary, Dhital et al. (2010).
\end{flushleft}
\end{table*}




\end{document}